\documentclass[%
reprint,
amsmath,amssymb,
aps,hidelinks,
]{revtex4-1}
\setcitestyle{super}
\usepackage{color}
\newcommand{\rev}[1]{\textcolor{black}{#1}}

\usepackage{hyperref}
\usepackage{graphicx}
\usepackage{dcolumn}
\usepackage{bm}
\usepackage{color}

\newcommand{\beginsupplement}{%
	\setcounter{table}{0}
	\renewcommand{\thetable}{S\arabic{table}}%
	\setcounter{figure}{0}
	\renewcommand{\thefigure}{S\arabic{figure}}%
}

\begin{document}
	
	\preprint{APS/123-QED}
	
	\title{Temporal sequences of brain activity at rest are constrained by white matter structure and modulated by cognitive demands}
	\author{Eli J. Cornblath$^{1,2}$}
	\author{Arian Ashourvan$^{2}$}
	\author{Jason Z. Kim$^{2}$}
	\author{Richard F. Betzel$^2$}
	\author{Rastko Ciric$^{3}$}
	\author{Azeez Adebimpe$^{3}$}
	\author{Graham L. Baum$^{1,2,3}$}
	\author{Xiaosong He$^{2}$}
	\author{Kosha Ruparel$^{3}$}
	\author{Tyler M. Moore$^{3}$}
	\author{Ruben C. Gur$^{3,6,8}$}
	\author{Raquel E. Gur$^{3,6,8}$}
	\author{Russell T. Shinohara$^{4}$}
	\author{David R. Roalf$^{3}$}
	\author{Theodore D. Satterthwaite$^{3}$}
	\author{Danielle S. Bassett$^{2,3,5,6,7,9,10}$}
	
	\affiliation{}
	\affiliation{
		$^1$Department of Neuroscience, Perelman School of Medicine
	}
	\affiliation{
		$^2$Department of Bioengineering, School of Engineering \& Applied Science
	}
	
	\affiliation{
		$^3$Department of Psychiatry, Perelman School of Medicine
	}
	\affiliation{
		$^4$Department of Biostatistics, Epidemiology, \& Informatics, Perelman School of Medicine
	}
	\affiliation{
		$^5$Department of Physics \& Astronomy, College of Arts \& Sciences
	}
	\affiliation{
		$^6$Department of Neurology, Perelman School of Medicine
	}
	\affiliation{
		$^7$Department of Electrical \& Systems Engineering, School of Engineering \& Applied Science 
	}
	\affiliation{
		$^8$Department of Radiology, Perelman School of Medicine, University of Pennsylvania, Philadelphia, PA, 19104 USA,
	}
	\affiliation{
		$^9$Santa Fe Institute, Santa Fe, NM, 87501 USA,
	}
	\affiliation{
		$^{10}$To whom correspondence should be addressed: dsb@seas.upenn.edu
	}
	
	\begin{abstract}	
		
		A diverse white matter network and finely tuned neuronal membrane properties allow the brain to transition seamlessly between cognitive states. However, it remains unclear how static structural connections guide the temporal progression of large-scale brain activity patterns in different cognitive states. Here, we analyze the brain's trajectories through a high-dimensional activity space at the level of single time point activity patterns from functional magnetic resonance imaging data acquired during passive visual fixation (rest) and an n-back working memory task. We find that specific state space trajectories, which represent temporal sequences of brain activity, are modulated by cognitive load and related to task performance. Using diffusion-weighted imaging acquired from the same subjects, we use tools from network control theory to show that linear spread of activity along white matter connections constrains the brain's state space trajectories at rest. Additionally, accounting for stimulus-driven visual inputs explains the different trajectories taken during the n-back task. We also used models of network rewiring to show that these findings are the result of non-trivial geometric and topological properties of white matter architecture. Finally, we examine associations between age and time-resolved brain state dynamics, revealing new insights into functional changes in the default mode and executive control networks. Overall, these results elucidate the structural underpinnings of cognitively and developmentally relevant spatiotemporal brain dynamics.
		
	\end{abstract}
	\maketitle
	\section*{Introduction}
	
	\rev{An elusive goal of computational neuroscience is to describe the brain as a dynamical system with a predictable natural temporal evolution and response to input. Such a model would be invaluable to clinicians as a generalizable tool for identifying optimal brain stimulation approaches to drive the brain from various states of disease to states of health \cite{Deco2014,Deco2019}. Yet, the endeavor of identifying a real non-linear dynamical system that provides such insights is exceedingly difficult, in part due to the high dimensionality of brain activity and the complex nature of the brain's intrinsic functional interactions. It is known that the white matter architecture of the brain contributes to the diverse patterns of activity and functional connectivity that represent information processing underlying cognitive function \cite{Honey2009,Cabral2017,Fukushima2017}. However, the exact manner in which white matter connectivity constrains the temporal dynamics of brain activity remains poorly understood. Improving our understanding requires a rich characterization of time-varying brain activity, as well as a robust model to link brain structure with brain activity.}
	
\begin{figure}
		\begin{center}
			\includegraphics[width=9cm,keepaspectratio]{\string 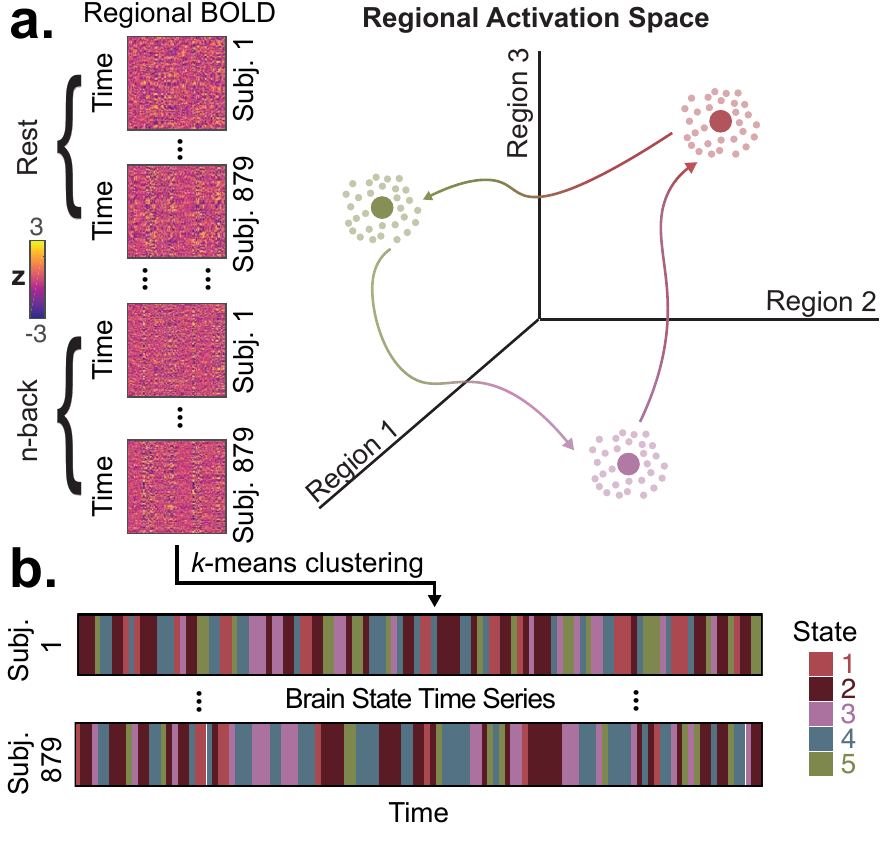}
			\caption{\textbf{Schematic of methods for functional image analysis.} \emph{(a)} Regional BOLD time series from resting-state and n-back task scans are concatenated across subjects. \rev{Each row in this concatenated data matrix represents a point in a high-dimensional space whose axes correspond to regional activity. A schematic of a low-dimensional version of this space is shown as an example on the right. Our goal is to identify frequently visited locations in this space and study the temporal progression between these locations during rest and task. }\emph{(b)} We then apply a $k$-means clustering algorithm to generate a series of cluster labels that can be mapped back to individual subjects, producing subject-specific brain state time series.
				\label{fig:figure1}}
		\end{center}
		
	\end{figure}

	\rev{Myriad approaches have been applied to resting functional magnetic resonance imaging (fMRI) to understand intrinsic brain dynamics. The most common approach (``functional connectivity'') involves analyzing the correlations between the activity time series of pairs of brain regions.} While pairwise correlation-based approaches summarize inter-regional synchrony over a period of time, cutting-edge signal-processing approaches to fMRI can provide a richer account of brain dynamics by considering the whole-brain patterns of activity at single time points \cite{Karahanolu2015,Chen2015,Liu2013,Vidaurre2017a,Taghia2018,Chen2018,Medaglia2018,Reddy2018}. One can conceive of the brain as progressing through a state space whose axes correspond to the activity at each region \cite{Shine2019,Saggar2018} (Fig. \ref{fig:figure1}a). Each point in this space corresponds to an observed pattern of brain activity, and the sequential trajectories through this space represent how brain activity patterns change over time. This approach allows one to utilize the maximum temporal resolution offered by BOLD fMRI, unlike many dynamic functional connectivity methods, which are limited by a minimum window size \cite{Preti2017}. Studies analyzing the brain's regional activation space have found that frequently visited activity patterns consist of different combinations of RSN components \cite{Vidaurre2017a,Karahanolu2015,Liu2013,Petridou2013,Tagliazucchi2012}. Brain activity patterns are known to represent information content \cite{Naselaris2011}, distinct modes of information processing \cite{Barch2013,Vossel2014}, and attention to stimuli \cite{Vossel2014,Gandhi1999}. Such activation patterns occur both at rest and in the presence of tasks or attentional demands, and are often considered to be neural representations of cognitive state \cite{Karahanoglu2017,Shine2019,Saggar2018}.
	
	However, a fundamental understanding of the brain's trajectories through regional activation space has been limited by the use of thresholding that disrupts the continuity of the time series \cite{Karahanolu2015,Chen2015,Liu2013}, a focus on between- rather than within-scan differences \cite{Saggar2018}, a narrow focus on only a few brain regions \cite{Taghia2018}, and various modeling assumptions impacting the nature of the temporal dynamics detected \cite{Vidaurre2017a}. Such limitations have also hampered progress in understanding how state-space trajectories might be constrained by or indeed supported by underlying brain structure. One intriguing possibility is that the white matter architecture of the brain is designed to support coordinated activity within RSNs and information transfer between RSNs, which might be reflected in the temporal progression between distinct states of RSN coactivation. For instance, one could imagine that coactivation of visual regions with dorsal attention regions, followed by activation of frontoparietal executive control regions might reflect reception, integration, and higher order processing of a visual stimulus. Critically, the normative neurodevelopment of time-resolved brain state dynamics and their cognitive relevance also remain unknown, limiting our ability to incorporate such neurobiological features into our understanding of neuropsychiatric disorders with developmental origins \cite{Thompson2014,Satterthwaite2016,Bassett2018}. Specific neuropsychiatric symptoms, such as hallucinations or negative rumination, may be represented in coactivation patterns and their temporal dynamics, which could be disrupted with brain stimulation \cite{Stiso2018,Silvanto2008,Rachid2018,Chen2013}.

	To address these fundamental gaps in knowledge, we consider a large, community-based sample ($n=879$) of healthy youth from the Philadelphia Neurodevelopmental Cohort \cite{Satterthwaite2014,Satterthwaite2016}, all of whom underwent diffusion- and T1-weighted structural imaging, passive fixation resting state fMRI, and n-back working memory task fMRI \cite{Ciric2017,Roalf2016,Rosen2018}. We begin by using $k$-means clustering to extract a set of discrete brain states from the fMRI data \cite{Chen2015,Liu2013,Chen2018,Gutierrez-Barragan2019}, and to assign each functional volume from both rest and task scans to one of those states. We hypothesize that the brain's temporal progression between different states is influenced by cognitive demands and stimuli, which we test by quantifying the time that subjects dwell within states, and the propensity to transition between states. Next, we hypothesize that structural connectivity constrains the temporal progression of brain states and explains why these particular brain states exist. We test these hypotheses using emerging tools from network control theory \cite{Gu2015,Gu2017,Betzel2016,Tang2017,Stiso2018}, along with comparison to stringent null models \cite{Alexander-Bloch2018,Betzel2018} to ensure the specificity of our findings. Finally, we hypothesize that brain state dynamics change throughout development to optimize cognitive performance.
	
	By rigorously testing these hypotheses, we find increased temporal persistence of a state associated with high activity in frontoparietal cortex during task. On the other hand, states associated with coherent activity in default mode areas have similar temporal persistence between rest and task with an increased rate of appearance during rest. \rev{Interestingly, two divergent trajectories towards frontoparietal and default mode states following from a sensory-driven state are positively and negatively related to task performance, respectively. Using tools from linear network control theory, we show that state transitions with small energy requirements given the brain's white matter architecture occur more frequently in the observed data than state transitions with large energy requirements. Additionally, accounting for visual input explains the differences in state-space trajectories between rest and task.} Finally, we show that brain state dynamics and predicted energies of state transitions are associated with age and explain individual differences in working memory performance. Overall, we demonstrate the utility of state-space models in understanding the structural basis for developmentally and cognitively relevant context-dependent brain dynamics.
	
	\section*{Results}
	
	\subsection*{Brain states capture instantaneous coactivation between resting state functional networks}
	
	The spatiotemporal dynamics of brain activity are exceedingly complex and not fully understood. Analyzing pairwise correlations between regions over time (``functional connectivity'' or FC) is a common approach used to quantify interactions between brain regions. However, static FC does not necessarily account for spontaneous or stimulus-evoked coactivation observed at single time frames (Fig. \ref{fig:figureS11}), which is the maximum temporal resolution offered by BOLD fMRI for a given repetition time (TR) \cite{Liu2013,Liu2013c}. Here, we used $k$-means clustering \cite{Liu2013,Gutierrez-Barragan2019,Cabral2017} to assign each time point from resting and n-back task fMRI scans into clusters of statistically similar and temporally recurrent whole-brain spatial coactivation patterns, hereafter referred to as ``brain states'' (Fig. \ref{fig:figure1}a). Importantly, we found that these BOLD data exhibited clustering in regional activation space beyond what would be expected from signals with the same autocorrelation profiles (Fig. \ref{fig:figureS13}a), and states were similar between rest and n-back task scans (Fig. \ref{fig:figureS2}b).
	
\begin{figure*}
		\begin{center}
			\includegraphics[width=18cm,keepaspectratio]{\string 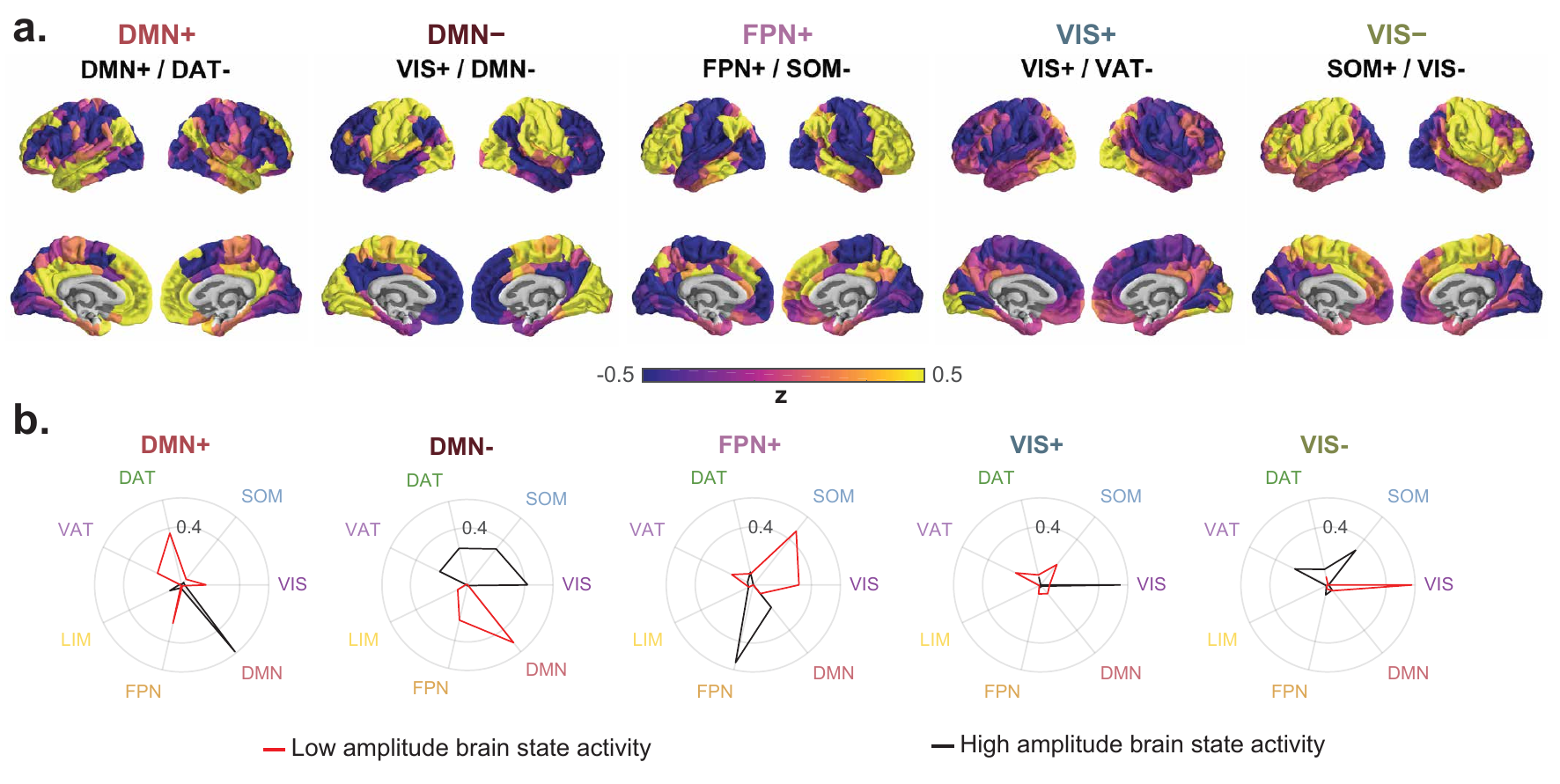}
			\caption{\textbf{Brain states represent coactivation within and between resting state functional networks.} \emph{(a)} Brain states defined as the centroids of clusters identified using an unsupervised machine learning algorithm applied to rest and n-back task fMRI data. Brain states are labeled based on cosine similarity with \textit{a priori} resting state functional networks (RSNs) \cite{ThomasYeo2011}. The top label corresponds to the RSN with the most overall similarity, and the bottom two labels separated by a forward slash reflect the RSNs with the most similarity to the positive and negative components of each state, respectively. \emph{(b)} Cosine similarity between positive (black) and negative (red) components of each state with binary state vectors corresponding to \textit{a priori} definitions of RSNs \cite{ThomasYeo2011}. Larger radial values correspond to higher cosine similarity. \emph{DAT}, dorsal attention network, \emph{DMN}, default mode network, \emph{FPN}, frontoparietal network, \emph{LIM}, limbic network, \emph{SOM} somatomotor network, \emph{VAT}, ventral attention network, and \emph{VIS}, visual network.
				\label{fig:figure2}}
		\end{center}
	\end{figure*}

	We found that ``resting-state functional networks'' (RSNs) \cite{ThomasYeo2011,Schaefer2017,Power2011}, groups of regions with stronger static FC with each other than with other regions, exhibited coherent high or low amplitude activity within each cluster centroid. This finding is consistent with strong \textit{within-network FC}. Due to their similarity to RSNs, we named each of the five states that we observed after the previous RSN whose isolated high or low amplitude activity best explained each state. This choice did not influence any analyses and is solely for convenient interpretation. We refer to them as the DMN+, DMN-, FPN+, VIS+, and VIS-, representing activity above (+) or below (-) regional means in default mode (DMN), frontoparietal (FPN), and visual networks (VIS), respectively (Fig. \ref{fig:figure2}a). We also asked which additional RSNs exhibited coherent activity in each state by quantifying the alignment of the high and low amplitude components of each brain state activity pattern separately with each RSN, indicating the presence of coherent activity within SOM (somatomotor network), DAT (dorsal attention network), and VAT (ventral attention network) (Fig. \ref{fig:figure2}b). 
	
	Interestingly, in addition to coherent activity within each RSN, we found that centroids contained multiple RSNs simultaneously exhibiting coherent high or low amplitude activity. For example, the DMN exhibited high amplitude while the DAT simultaneously exhibited low amplitude in the DMN+ state. This spatial organization likely reflects known patterns of \textit{between-network FC} between task-positive and task-negative systems \cite{Fox2005} (Fig. \ref{fig:figureS12}a, mean $r=-0.10$, one-sample $t$-test, $df=878$, $t=-80.45$, $p<10^{-15}$). However, the DMN-, VIS+, and VIS- states evidence unexpected, transient patterns of coactivation between VIS and SOM systems. Specifically, in the DMN- state, SOM and VIS regions were both at high amplitude (Fig. \ref{fig:figure2}b). In the VIS+ state, SOM regions were at low amplitude and VIS regions were at high amplitude (Fig. \ref{fig:figure2}b). In the VIS- state, SOM regions were at high amplitude and VIS regions were at low amplitude (Fig. \ref{fig:figure2}b). Despite the presence of three unique coactivation patterns between these two RSNs, the mean FC between regions in VIS and SOM did not significantly differ from 0 (Fig. \ref{fig:figureS12}a, mean $r=-0.0012$, one-sample $t$-test, $df=878$, $t=-0.085$, $p=0.40$). These patterns of simultaneous activation and deactivation provide a snapshot of instantaneous interactions between RSNs that could not be obtained through the analysis of FC. 
	
	\subsection*{Temporal patterns of brain state occurrence and occupancy}
	
\begin{figure*}
		\begin{center}
			\includegraphics[width=18cm,keepaspectratio]{\string 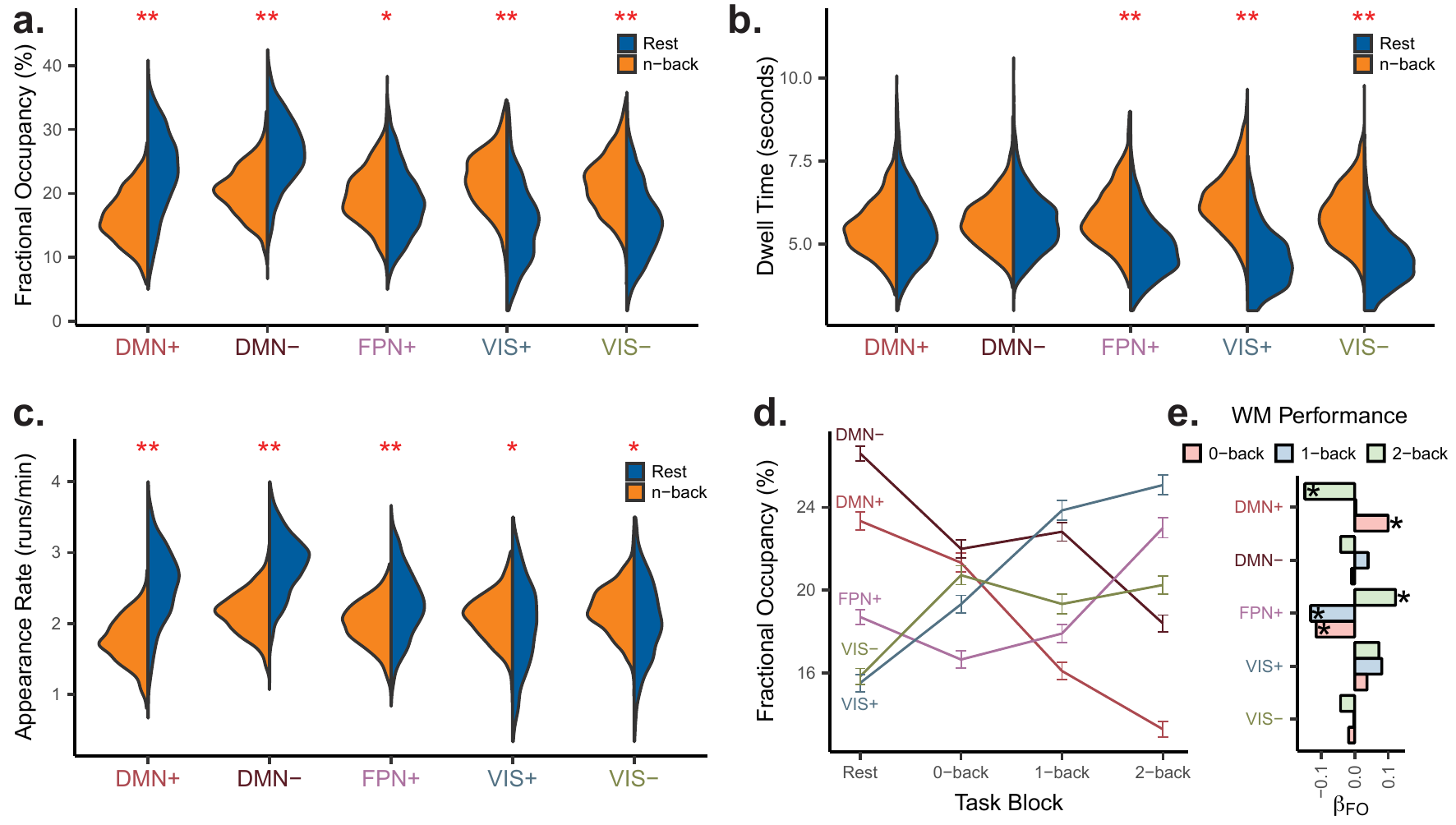}
			\caption{\textbf{Brain state occupancy and temporal persistence are modulated by task demands.} \emph{(a, b, c)} Distributions of subject-level fractional occupancy (panel \emph{a}), dwell time (panel \emph{b}), and appearance rate (panel \emph{c}), for each brain state in rest and task. DMN states exhibit higher fractional occupancies and appearance rates during rest and VIS states exhibit higher fractional occupancies and dwell times during task. **, $p_\mathrm{corr} < 10^{-15}$, *, $p_\mathrm{corr} < 10^{-4}$, paired $t$-tests Bonferroni-corrected across $k=5$ states separately for fractional occupancy, dwell time, and run rate. \emph{(d)} Within task scans, the state dwell times change with increasing cognitive load. The thick colored lines indicate the mean across subjects, and the error bars represent 2 standard errors above and below the mean for each block. \emph{(e)} Standardized linear regression $\beta$ weights for state-specific fractional occupancy (FO) on working memory (WM) performance for each task block requiring an increasing WM load (0-back, 1-back, and 2-back). We found opposing trends for DMN+ and FPN+ states from 0-back to 2-back. \emph{WM}, working memory. \emph{FO}, fractional occupancy.
				\label{fig:figure2_5}}
		\end{center}
	\end{figure*}

	After identifying large-scale brain states representing instantaneous coactivation between RSNs, we were interested in comparing the dynamics of brain state occupancy and dwelling between rest and n-back scans (Fig. \ref{fig:figure2_5}a). To provide a rich characterization of the dynamics of brain state occupancy, we defined and studied three related metrics for each state: (1) \textit{fractional occupancy}, the percentage of frames assigned to a state for a given scan or condition, (2) \textit{dwell time}, the mean duration in seconds of temporally continuous runs of state occupancy, and (3) \textit{appearance rate}, the number of times a run of any length appeared per minute. Using paired $t$-tests, we assessed whether the population means of subject-specific differences between n-back and rest ($\mu_{nback-rest}$) for each of these metrics were different from 0. Here, we focus on the FPN and DMN, whose activation and suppression, respectively, are classically seen during cognitively demanding tasks \cite{Anticevic2012,Fox2005,Scolari2015}.
	
	During the n-back task, we observed lower fractional occupancies in the two default mode states (paired $t$-tests, $df=878$, $t=-31.38$, DMN+: $\mu_{nback - rest} = -7.10$, $p_\mathrm{corr} < 10^{-15}$, DMN-: $\mu_{nback - rest} = -6.15$, $df=878$, $t=-30.57$, $p_\mathrm{corr} < 10^{-15}$). However, higher fractional occupancy in DMN states at rest was best explained by increased appearance probability of DMN states at rest (paired $t$-tests, DMN+: $\mu_{nback - rest} = -0.77$, $df=878$, $t= -40.03$, $p_\mathrm{corr} < 10^{-15}$, DMN-: $\mu_{nback - rest} = -0.68$, $df=878$, $t=-40.74$, $p_\mathrm{corr} < 10^{-15}$), while dwell time in DMN states did not differ between rest and task (paired $t$-tests, DMN+: $\mu_{nback - rest} = -0.01$, $df=878$, $t= -0.17$, $p_\mathrm{corr} = 1$, DMN-: $\mu_{nback - rest} = 0.06$, $df=878$, $t=1.45$, $p_\mathrm{corr} = 0.73$). Lower DMN+ state fractional occupancies during the n-back task is consistent with DMN suppression observed during attention-demanding tasks \cite{Anticevic2012}. However, the high DMN- fractional occupancy suggests that coherent DMN suppression is not specific to task conditions, and may occur in the context of a unique, transient interaction with primary sensory areas (Fig. \ref{fig:figure2_5}a). Interestingly, FPN+ state fractional occupancy was similar between rest and task, despite higher dwell time with a lower appearance rate in the n-back task (paired $t$-tests, FPN+ dwell time: $\mu_{nback - rest} = 0.95$, $df=878$, $t= 23.69$, $p_\mathrm{corr} < 10^{-15}$, FPN+ appearance rate: $\mu_{nback - rest} = -0.26$, $df=878$, $t=-14.74$, $p_\mathrm{corr} < 10^{-15}$). These findings suggest that the FPN is activated more frequently albeit transiently at rest, while sustained activation of the FPN is found during the n-back working memory task.
	
	Next, we decided to examine the dynamics of DMN suppression and FPN activation as a function of cognitive load within the n-back task and as a predictor of task performance. We hypothesized that as cognitive load increased, DMN+ fractional occupancy would decrease and FPN+ fractional occupancy would increase. As expected, the FPN state fractional occupancy increased from the 0-back to the 2-back block (Fig. \ref{fig:figure2_5}d). Interestingly, spatially anticorrelated DMN states both decreased with increasing cognitive load (Fig. \ref{fig:figure2_5}d). This finding suggests that working memory involves reduced representation of brain states with coherent activity in the DMN, whether high or low amplitude, and increased representation of the high amplitude FPN state, clarifying the roles of task-positive and task-negative networks \cite{Anticevic2012,Fox2005,Scolari2015}. Next, when we examined associations between fractional occupancy and block-specific working memory performance (Fig. \ref{fig:figure2}c-d), we found that increasing FPN+ fractional occupancy (Fig. \ref{fig:figure6}c; multiple linear regression, standardized $\beta_{FO} = 0.12$, $df=872$, $t=3.85$, $p_\mathrm{corr} = 1.9 \times 10^{-3}$) and decreasing DMN+ fractional occupancy (Fig. \ref{fig:figure6}c; multiple linear regression, standardized $\beta_{FO} = -0.15$, $df=872$, $t=-4.71$, $p_\mathrm{corr} =  4.4\times 10^{-5}$) were associated with working memory performance during the 2-back block. However, for 0-back blocks, these trends were reversed (Fig. \ref{fig:figure6}c, multiple linear regression; 0-back FPN+, standardized $\beta_{FO} = -0.11$, $df=872$, $t=-3.49$, $p_\mathrm{corr} = 7.7 \times 10^{-3}$; 0-back DMN+, standardized $\beta_{FO} = 0.10$, $df=872$, $t=2.96$, $p_\mathrm{corr} = 0.047$). This pattern of results might reflect the engagement of alternative systems for low difficulty tasks by strong performers, thus introducing a layer of complexity to the notion of DMN and FPN as primary task-negative and task-positive systems \cite{Fox2005}.
	
	\subsection*{Transitions between brain states}
	
	After demonstrating that cognitive demands influence dwell times \textit{in} large-scale brain states, we were interested in how cognitive demands would affect transitions \textit{between} large-scale brain states. We conceptualized brain state transitions as directional trajectories between different locations in a high-dimensional space whose axes correspond to the level of activity in each brain region. Neuroimaging studies suggest that the brain progresses along a low-dimensional manifold in regional activation space \cite{Shine2019,Saggar2018}, but it remains unknown the extent to which specific trajectories in this space are influenced by cognitive demands and may represent cognitive processes. 
	
\begin{figure*}
		\begin{center}
			\includegraphics[width=18cm,keepaspectratio]{\string 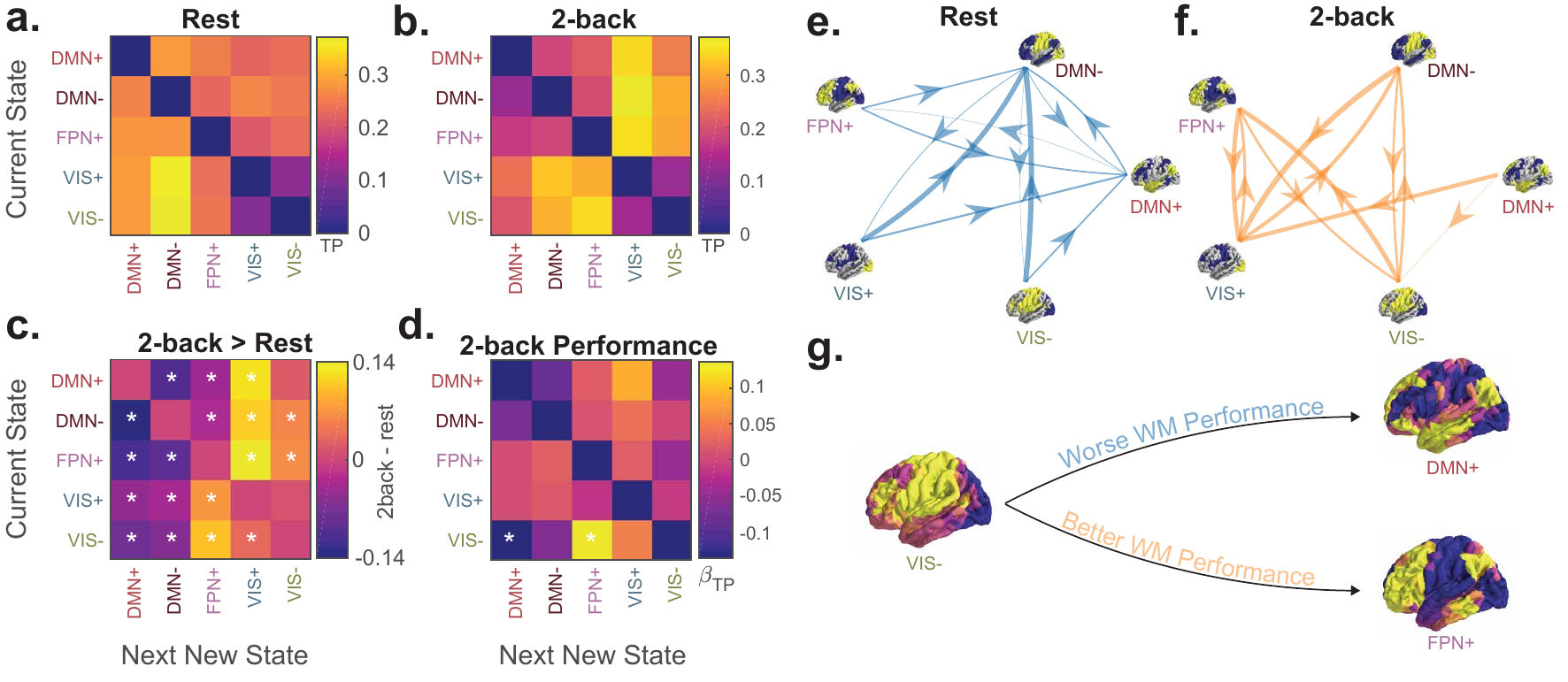}
			\caption{\textbf{Brain state transitions are influenced by task demands and related to behavior.} \emph{(a-b)} Group average state transition probability matrices for resting state scans and the 2-back condition. Matrix elements reflect the probability of a state transition after removing the effects of state autocorrelation. \emph{(c)} Non-parametric permutation testing demonstrating differences between the rest and n-back group average transition probability matrices. *, $p_\mathrm{corr} < 0.05$, after Bonferroni correction over 20 transition probabilities. \rev{\emph{(d)} Standardized linear regression $\beta$ weights for the transition probability during the 2-back condition of the n-back task as a predictor of task performance during the 2-back condition. Transitions from the VIS- state into the DMN+ and FPN+ states are negatively and positively associated with better performance, respectively. *, $p_\mathrm{corr} < 0.05$, after Bonferroni correction over 20 transition probabilities. \emph{TP}, transition probability. \emph{(e-f)} Graphical representation of resting state (panel \emph{e}) and 2-back (panel \emph{f}) transition probability matrices as networks whose nodes are states, and whose edges are transition probabilities thresholded at 0.25. \emph{(g)} Graphical representation of results shown in panel \emph{d}.}
				\label{fig:figure3}}
		\end{center}
	\end{figure*}

	\rev{Here, in order to study the relationship between cognition and progression through regional activation space, we computed transition matrices for each subject's resting state scan, n-back task scan, and each condition of the n-back task scan. Because we were interested in state changes, we constructed transition matrices that ignore the potentially independent effects of state persistence, or autocorrelation, and only capture the probabilities of moving to new states; that is, the $ij^{\mathrm{th}}$ element of the transition matrix represents the transition probability between state $i$ and state $j$ given that a transition out of state $i$ is occurring (Fig. \ref{fig:figure3}a, see Methods). As an initial step, we used two null models to confirm previous findings \cite{Vidaurre2017} that brain state transitions are non-random, in that the observed transition probabilities would be unlikely in uniformly random sequences of states and state transitions (Fig. \ref{fig:figureS13_5}).}
	
	Next, we explored how cognitive load impacts brain state transitions using a non-parametric permutation test to assay for differences between transition matrices computed from resting state scans and from the 2-back condition of the n-back task. \rev{We hypothesized that we would see more transitions from states driven by sensory cortex activation into states driven by activation in executive control and attention areas, reflecting reception, integration, and task-relevant processing of stimuli. Indeed, we found that transitions from VIS+ and VIS- states into the FPN+ state were increased during the 2-back condition compared to rest scans (Fig. \ref{fig:figure3}b). Transitions from DMN+, DMN-, and FPN+ states into VIS+ states were also increased in the 2-back condition, likely reflecting the interruption of ongoing transmodal processing by sensory input. Finally, we tested for associations between 2-back transition probabilities and performance during the 2-back condition. In support of our hypothesis, we found that transitions from the VIS- state to the DMN+ state were negatively associated with performance (Fig. \ref{fig:figure3}d, multiple linear regression, standardized $\beta_{TP} = -0.14$, $df=873$, $t=-4.42$, $p_\mathrm{corr} = 2.24 \times 10^{-4}$), while transitions from the VIS- state to the FPN+ state were positively associated with performance (Fig. \ref{fig:figure3}d, multiple linear regression, standardized $\beta_{TP} = 0.14$, $df=873$, $t=4.37$, $p_\mathrm{corr} = 2.84 \times 10^{-4}$). These results are consistent with prior work positing roles for the FPN and DMN as task-positive and task-negative systems \cite{Fox2005}, but suggest that interactions with motor, visual, and salience networks found in the VIS- state may also contribute to working memory. Overall, these findings suggest that specific trajectories in brain activation space are favored during increased cognitive load and may represent task-relevant processing.}
	
	\subsection*{Control properties of white matter networks explain brain state transitions}
	
\begin{figure*}
		\begin{center}
			\includegraphics[width=18cm,keepaspectratio]{\string 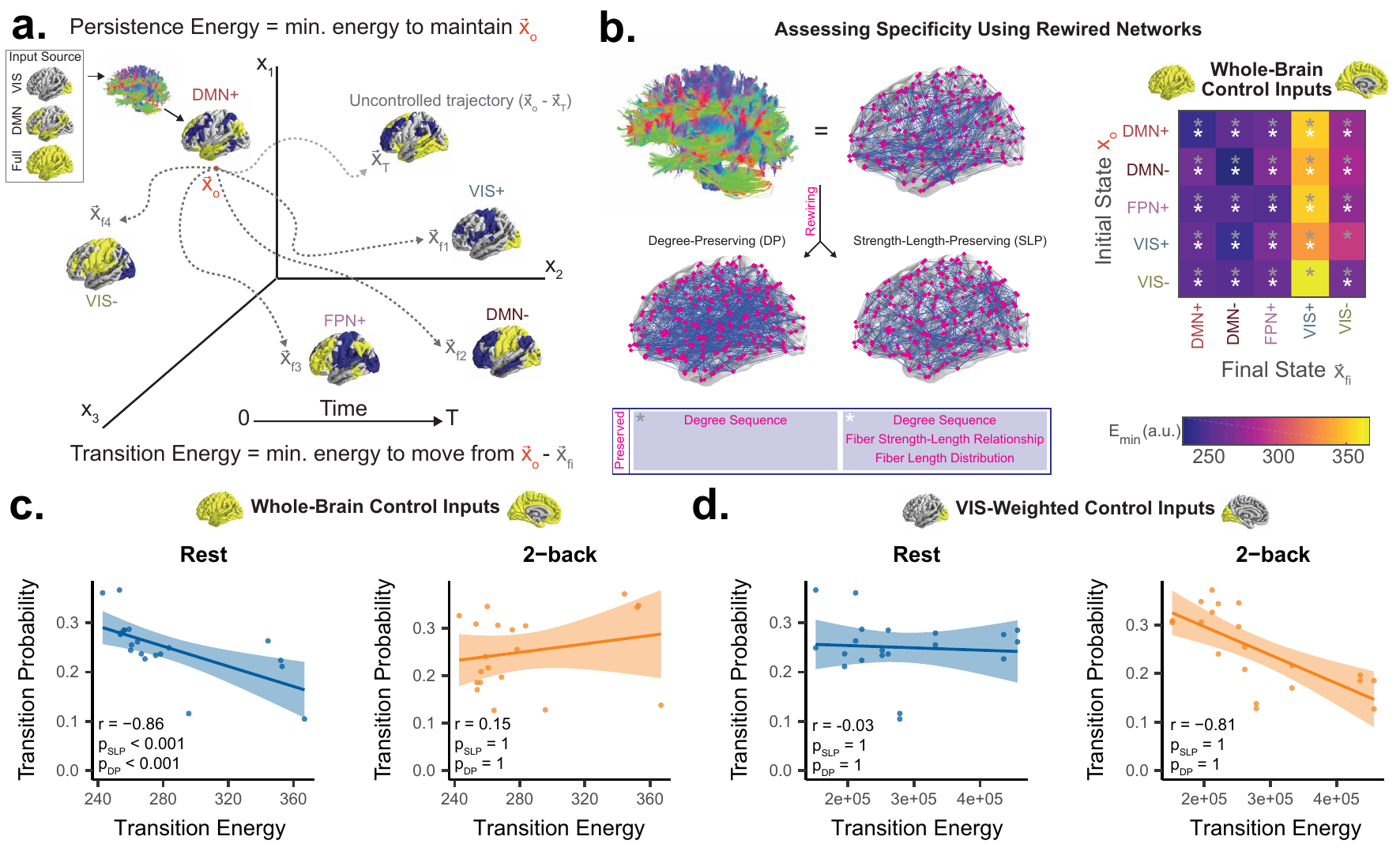}
			\rev{\caption{\textbf{Linear dynamics along white matter explain brain state transition probabilities.} \emph{(a)} Schematic demonstrating calculation of minimum control energy needed to move a linear dynamical system defined by white matter connectivity from some initial state $\mathbf{x}_o$ to some final state $\mathbf{x}_{f_i}$ over a time horizon $T$. \emph{(b)} Schematic of network null models (left) preserving different spatial and topologic features of networks defined by white matter connectivity. The energies (E\textsubscript{min}) required to maintain or transition between each state are lower in real brain networks compared to these null models (right). \emph{(c-d)} Correlation between structure-based transition energy prediction ($x$-axis) and empirically derived transition probability ($y$-axis) for resting state (left) and the 2-back condition of the n-back task (right), using inputs weighted evenly throughout the whole brain \emph{(c)} or weighted positively towards the visual system \emph{(d)}.
				\label{fig:figure5}}}
		\end{center}
	\end{figure*}

	In the previous section, we described how the presence of cognitive demands and sensory inputs leads the brain towards certain trajectories in state space. However, it is not well understood how the static white matter connectome contributes to these divergent dynamics. Here, we modeled the influence of structure on brain activity as the time-evolving state of a linear dynamical system defined by white matter connectivity. By applying tools from network control theory (Fig. \ref{fig:figure5}a; see Methods, subsection ``Network Control Theory'' and Supplementary Information, subsection ``Calculating transition energy using control theory''), we calculated the \textit{transition energy} as the minimum input energy needed to transition between every pair of the empirically observed brain states. In all calculations, we allowed the inputs to come from all brain regions, weighted either uniformly or towards a particular cognitive system \cite{ThomasYeo2011}. Using this framework, we tested a series of hypotheses unified under the notion that the brain prefers trajectories through state space requiring minimal input energy given structural constraints.
	
	First, we hypothesized that the brain is optimized to support the observed brain states and state transitions with relatively little energy. We measured brain state stability as \textit{persistence energy}, or the energy needed to maintain each state. In a single representative human structural brain network \cite{Betzel2017,Misic2015,Roberts2017a} (see Methods for details), we compared the transition and persistence energies for real structural connectivity (Fig. \ref{fig:figure5}b) and for two null models based on the group average human structural brain network: (1) a null model that preserves only degree sequence in the networks \cite{Rubinov2010} (Deg. Pres., DP), and (2) a null model that preserves degree sequence, edge length distribution, edge weight distribution, and edge weight-length relationship \cite{Betzel2018} (Strength-Length Preserving, SLP). Compared to the DP null model, transition and persistence energy were always lower in the group average SC (Fig. \ref{fig:figure5}b, all $p_\mathrm{corr}<0.001$). \rev{Compared to the SLP null model, every single persistence energy value and all but two transition energy values were lower in the group average SC (Fig. \ref{fig:figure5}b, $p_\mathrm{corr}<0.001$). Finally, we found that the energy required to maintain the DMN+ state was lower than a set of null states with similar spatial covariance \cite{Alexander-Bloch2018} (Fig. \ref{fig:figureS19}).} Collectively, these findings suggest that unique geometric and topological features of white matter networks allow for low energy transitions and maintenance of the empirically observed functional states.
	
	\rev{Next, we hypothesized that the brain prefers trajectories through state space that require little input energy to achieve in a dynamical system defined by white matter connectivity. To test this hypothesis, we computed the Spearman correlation between transition energy values and transition probabilities observed during resting state scans and during the 2-back condition of n-back task scans (Fig. \ref{fig:figure5}c-d). When inputs are evenly weighted throughout the whole brain (Fig. \ref{fig:figure5}c), transition energy values are strongly anticorrelated with resting state transition probabilities and weakly correlated with 2-back transition probabilities. Importantly, the energy estimates from real structural connectivity were more strongly anticorrelated with resting state transition probabilities than energy estimates from null models or transition distance in state space alone (Spearman's $r=-0.86$, $p_{SLP} < 0.001$, $p_{DP}<0.001$, Fig. \ref{fig:figureS20}b). When inputs are biased towards the visual system \cite{ThomasYeo2011} (Fig. \ref{fig:figure5}d), transition energy values are strongly anticorrelated with 2-back transition probabilities (Spearman's $r=-0.81$) and weakly correlated with resting state transition probabilities (Spearman's $r=-0.03$). However, this result was primarily explained by transition distance in state space, rather than the effects of structure ($p_{SLP} = 1$, $p_{DP} = 1$). Overall, these findings suggest that linear diffusion of brain activity along white matter tracts constrains brain state transitions at rest, and that the distribution of inputs to the brain is an important factor in the brain's progression through state space.}
	
	\subsection*{Brain state dynamics and control energies are associated with age}
	
	\begin{figure*}
		\begin{center}
			\includegraphics[width=18cm,keepaspectratio]{\string 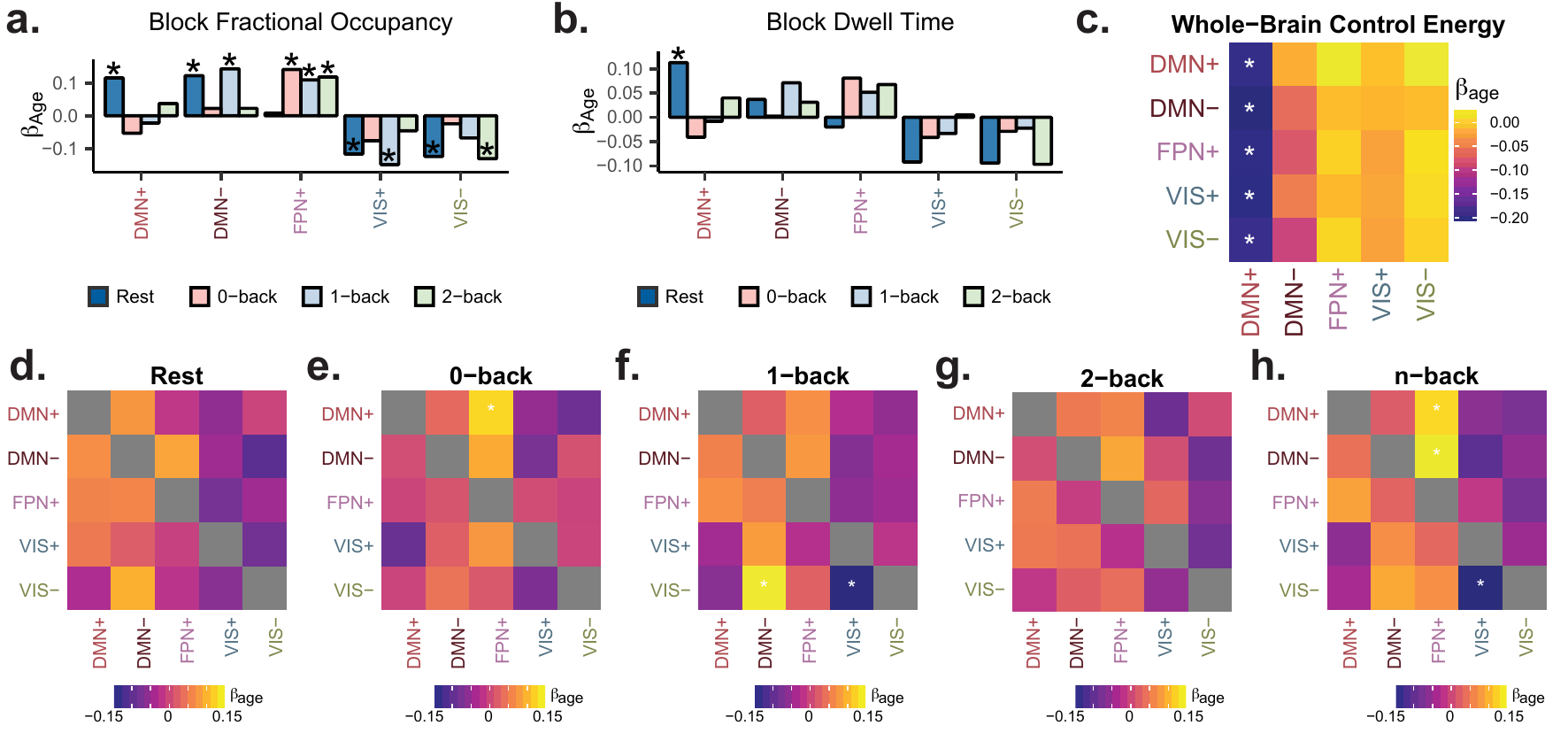}
			\caption{\textbf{Brain state dynamics and control energies are associated with age.} \emph{(a-b} Standardized linear regression $\beta$ weights for age as a predictor of fractional occupancy \emph{(a)} or dwell time \emph{(b)} in each state during rest and during each condition of the n-back task. *, $p_\mathrm{corr} < 0.05$ after Bonferroni correction over 20 state transitions. \rev{\emph{(c)} Standardized linear regression $\beta$ weights for age as a predictor of minimum control energy required to transition between each pair of states. *, $p_\mathrm{corr} < 0.05$ after Bonferroni correction over 20 state transitions.} \emph{(d-h)} Standardized linear regression $\beta$ weights for age as a predictor of transition probabilities during resting state scans \emph{(d)}, \rev{ 0-back \emph{(e)}, 1-back \emph{(f)}, and 2-back \emph{(g)}} conditions of the n-back task, and the entire n-back task scan \emph{(h)}. *, $p_\mathrm{corr} < 0.05$ after Bonferroni correction over 100 state transitions shown in panels \emph{d-h}.
				\label{fig:figure6}}
		\end{center}
	\end{figure*}
	
	Developmental changes in white matter, grey matter, functional networks, and task-related activations accompany changes in behavior and cognition \cite{Richards2015,Power2010,Satterthwaite2013,Betzel2014,Hutchison2015}. However, it is unclear how state space trajectories and their supporting structural features contribute to these cognitive and behavioral changes. Given that the spatiotemporal brain dynamics identified by our approach have clear structural underpinnings, we hypothesized that these dynamics change throughout normative neurodevelopment in support of emerging cognitive abilities \cite{Tang2017a,Cui2018}. 
	
	We used multiple linear regression to ask whether age was associated with state dwell times and fractional occupancies while controlling for brain volume, handedness, head motion, and sex as potential confounders. Interestingly, we found that fractional occupancies in FPN+ and DMN+ states exhibited context-dependent associations with age (Fig. \ref{fig:figure6}a). FPN+ fractional occupancy increased with age for all blocks of the n-back task (Fig. \ref{fig:figure6}a; multiple linear regression, 2-back standardized $\beta_{age} = 0.12$, $df=873$, $t=3.40$, $p_\mathrm{corr} = 0.014$) and not rest, while DMN+ fractional occupancy increased with age for rest only (Fig. \ref{fig:figure6}a; multiple linear regression, standardized $\beta_{age} = 0.12$, $df=873$, $t=3.59$, $p_\mathrm{corr} = 0.015$). The relationships between dwell time and age followed similar but weaker trends to those observed with fractional occupancy, with the exception of resting state DMN+ state dwell time which increased with age. \rev{We also found that the minimum control energy required to undergo all transitions that terminated in the DMN+ state decreased with age (Fig. \ref{fig:figure6}c; all $p_{\mathrm{corr}} < 0.05$). This finding suggests that age-associated structural changes allow individuals to coherently activate the default mode network with greater ease, and is consistent with the observation that DMN+ dwell time and fractional occupancy increase with age at rest.}
	
	We also assessed whether transition probabilities were associated with age. Using multiple linear regression, we tested for relationships between transition probabilities or transition energy values and age, while controlling for brain volume, handedness, head motion, and sex. Similar to the context-dependent associations with age that we observed with fractional occupancy, we found that transition probabilities were differentially associated with age across the conditions of the n-back task (Fig. \ref{fig:figure6}e-h). The probability of transitions from both DMN- (Fig. \ref{fig:figure6}f; multiple linear regression, standardized $\beta_{age} = 0.15$, $df=873$, $t=4.33$, $p_\mathrm{corr} = 1.7 \times 10^{-3}$) and DMN+ (Fig. \ref{fig:figure6}f; standardized $\beta_{age} = 0.13$, $df=873$, $t=3.68$, $p_\mathrm{corr} = 0.025$) into FPN+ during the n-back task increased with age. This observation is particularly interesting in light of previous work implicating the DMN and FPN in increasing working memory performance across development \cite{Satterthwaite2013}. Specifically, this result provides evidence for the importance of direct switching between DMN and FPN states, as opposed to deactivation and activation without any temporal constraints. Overall, these findings suggest that task-oriented and spontaneous brain dynamics involving the DMN and FPN may mature through independent processes.
	
	\section*{Discussion}
	
	In the present study, we examined the temporal sequence of whole-brain activity patterns in individuals during rest and task, and demonstrate a structural basis for large-scale brain activity patterns and their dynamic temporal evolution. Using a diverse array of techniques from network neuroscience, dynamical systems, and network control theory, we generated new insights into the complex relationship between brain structure, spatiotemporal patterns of brain activity, neurodevelopment, and behavior.
	
	\subsection*{Time-resolved brain state dynamics}
	
	Cognitive functions are often represented as brain activity patterns \cite{Barch2013}, but substantially less is known about how sequences of activity patterns may represent links between cognitive functions. In this paper, we considered each fMRI image acquisition to be a point in a high-dimensional state space whose axes correspond to regional activity. Next, we identified brain states as frequently visited locations in this space comprised of combinations of active and inactive brain networks \cite{ThomasYeo2011}. Finally, we described the directional trajectories between these states in time as state transitions. Our work adds to a body of literature suggesting that coactivation of brain networks at relatively short temporal scales evidences rich functional interactions supporting behavior \cite{Vidaurre2017a,Karahanolu2015,Chen2018}. For instance, we found that the brain state transition probabilities observed at rest were strongly modulated by cognitive demand. During resting state scans, in which external stimuli are constant over time, transitions likely occur spontaneously, while during n-back task scans, transitions are likely caused by a combination of spontaneous fluctuations, stimulus-evoked activity in primary sensory areas, and task-related activity changes in higher order association areas. In the n-back task, we found more frequent transitions into states driven by coactivation of sensory systems from states involving coactivation of higher order association areas when compared to the resting-state. This finding was present in two independent samples (PNC and HCP) with different task structure and is consistent with increased top-down modulation \cite{Vossel2014} of sensory input during task performance. 
	
	\rev{We also found that certain trajectories in state space were related to task performance. The VIS- state occurred more frequently during the n-back task and is composed of visual cortex suppression alongside mixed dorsal and ventral attention network activation, consistent with top-down suppression of sensory cortex \cite{Vossel2014}. While its occupancy alone was not related to task performance, transitions from VIS- to FPN+ or DMN+ were positively and negatively associated with performance, respectively. These findings suggest that early stimulus processing followed by manipulation \cite{DEsposito1999} of task-relevant information facilitates accurate performance, while stimulus processing followed by internally directed cognition \cite{Anticevic2012} is detrimental to performance. More broadly, this result suggests that the paths through which activation patterns are reached are important, in addition to the activation patterns themselves.}

	\subsection*{Structural constraints on brain state transitions}
	
	Our major contribution to cognitive neuroscience and applied network science lies in describing how linear diffusion of activity along a static white matter architecture constrains trajectories through brain activation space at rest. We hypothesized that the state space of brain activity could be explained by two components: linear diffusion of activity along white matter tracts \cite{Abdelnour2014} and some nonlinear inputs, which include but are not limited to neuronal membrane dynamics, metabolic factors, and external stimuli. Under this model, we solved for the magnitude of these nonlinear inputs required to maintain and transition between brain states, given the constant constraint of linear diffusion of activity along white matter tracts.
	
	Using this approach, we found that the brain empirically prefers trajectories in state space requiring the least energy needed to overcome structural constraints for a given set of inputs. Specifically, when we modeled uniformly weighted inputs or input weighted towards the DMN, the resulting transition energy values best explained resting state transition probabilities, possibly reflecting a regime with heterogeneous drivers centered around the DMN \cite{Anticevic2012}. As expected, these transition energies did not explain transition probabilities during the 2-back condition, likely due to task-derived inputs which were not explicitly modeled. Indeed, when we weighted system input towards the visual system to account for the frequent delivery of visual stimuli, we were better able to explain 2-back transition probabilities. While this finding represents constraints of distance alone and not white matter topology, it quantitatively explains how that stimulus-derived input alters the state-space trajectories of the brain. Future investigations may resolve the effects of structure on task dynamics using data-driven approaches that attempt to recover the full set of task-related inputs \cite{Ashourvan2019,Becker2018}.

	\subsection*{Age-associated brain state dynamics}
	
	Unlike previous time point level fMRI analyses \cite{Vidaurre2017a,Karahanolu2015,Chen2018}, our method unambiguously labels every time point in every subject for rest and n-back as belonging to a discrete, common state. We intentionally designed our method in this way to make comparisons across contexts and across subjects throughout different developmental stages. Indeed, these comparisons revealed context-specific associations between age and brain state dynamics, suggesting that as brain structure develops, multiple trajectories through state space are supported. Our study offered insights into previously unexplored time-resolved brain dynamics in normative neurodevelopment. Neuropsychiatric illnesses such as schizophrenia, autism, epilepsy, and ADHD are increasingly considered developmental disorders, and therefore it is critical to understand the maturation of brain dynamics in healthy youth. Previous studies have shown that structural and functional changes in the DMN and FPN accompany normal cognitive development \cite{Supekar2010,Luna2010,Taki2013}. Here we contribute to our understanding of these networks by demonstrating context-dependent associations between age and DMN and FPN state dynamics (Fig. \ref{fig:figure6}a,d-e). Interestingly, both fractional occupancies and state transition probabilities exhibit context-dependent associations with age, with DMN+ fractional occupancies increasing with age at rest only (Fig. \ref{fig:figure6}a), FPN+ fractional occupancies increasing with age in n-back only (Fig. \ref{fig:figure6}a), and DMN to FPN+ transitions increasing with age during task only (Fig. \ref{fig:figure6}e). \rev{However, like other cross-sectional studies of the relationship between brain function and age \cite{Satterthwaite2013b,Sato2014,Satterthwaite2013}, we found relatively small effects of age on individual measures.} Consistent with the finding of DMN+ fractional occupancy increasing with age, we also found that the predicted energy of transitioning into the DMN+ state from all other states decreased with age. However, we did not find a significant relationship between DMN+ transition energy values and DMN+ fractional occupancy across subjects. Future work should explore how the relative architecture of control energies within subjects may explain a bias towards certain trajectories over others.
	
	\subsection*{Methodological limitations}
	We acknowledge that a limitation of this study was a focus on discrete brain states with common spatial activity patterns across subjects rather than continuously fluctuating, overlapping functional modes of brain activity \cite{Shine2018,Karahanolu2015}. However, this simplified approach also constituted a major strength of the study, because it allowed us to assess age associations and cognitive effects of previously unexplored brain dynamics across subjects in a large sample. Generating discrete states also allowed us to examine the brain using approaches from stochastic process theory \cite{kwakernaak1972linear,cox2017theory}, including calculating transition probabilities. Importantly, our approach inherently accounts for the temporal autocorrelation within the BOLD signal \cite{Woolrich2001} by measuring state transitions while excluding state persistence. We also demonstrated that $k=5$ yields stable cluster partitions robust to outliers (Fig. \ref{fig:figureS1}), and our results were consistent for multiple values of $k$ (Fig. \ref{fig:figureS7}).
	
	The relatively low sampling rate (TR = 3s) likely limited our ability to resolve fast changes in brain activity. Nevertheless, we were able to resolve the effects of specific brain state transitions on behavior (Fig. \ref{fig:figure3}d, g). Additionally, there likely exist meaningful differences in individual brain state topographies \cite{Kong2018,Chong2017,Gordon2017} that certainly warrant further investigation, but could not be studied convincingly here due to the relatively small number of time frames acquired for each subject. To partially address these limitations, we reproduced key findings in a second parcellation (Fig. \ref{fig:figureS8}) and an independent sample with a higher sampling rate and no global signal regression (Fig. \ref{fig:figureS3}).
	
	\subsection*{Future directions}
	
	The novel approaches in this study pave the way for many future studies to continue to elucidate how a static structural connectome can give rise to complex, time-evolving activity patterns important for cognition. An intuitive and important application of our approach lies in the field of neurostimulation, where clinicians aim to implement targeted changes in the temporal evolution of brain activity patterns \cite{Stiso2018,Muldoon2016a,Chen2013} to alleviate symptoms of neuropsychiatric illness. In particular, network control theory and data-driven estimation of brain states are a powerful combination for this purpose. However, before this application can be realized, the robustness of these models at the level of individual subjects must be confirmed. One could similarly ask whether individual differences in structural connectivity explain variance in brain state dynamics, and thus response to neural stimulation. Application of these methods to electrophysiologic data would help to validate the dynamics that we observed and elucidate more complex neural dynamics that are not reflected in the slow fluctuations of hemoglobin oxygenation captured by BOLD fMRI \cite{Mateo2017}. Nonlinear neural mass models are powerful tools for understanding brain activity, and future work should attempt to validate the input-dependent, structure-based energetic constraints on state space trajectories observed in this study.
	
	Targeted, model-informed brain stimulation \cite{Deco2019,Muldoon2016a,Stiso2018} will likely need to account for interactions between exogenous input and endogenous dynamics \cite{Ashourvan2019,Becker2018}. Recent evidence \cite{Shine2019} implicates ascending neuromodulatory inputs in the brain's progression through state space. Release of neuromodulators can be driven by external stimuli or spontaneous neural activity \cite{Avery2017}, and therefore may serve as both an important mediator of external inputs and a critical aspect of endogenous dynamics. Ultimately, a model that integrates the dynamic and static interactions between brain structure, neuromodulators, fast ionotropic neurotransmission, and exogenous inputs might allow clinicians to solve for inputs that effect beneficial changes in brain activity and connectivity. Non-linear neural mass models of brain activity hold significant promise for this purpose \cite{Deco2018,Breakspear2017,Muldoon2016a} and in the future should attempt to incorporate the constraints of structure-based linear dynamics identified here.

	\section*{Methods}
	\subsection*{Participants}
	
	Resting state functional magnetic resonance imaging (fMRI), n-back task fMRI, and diffusion tensor imaging (DTI) data were obtained from $n=1601$ youth who participated in a large community-based study of brain development, known as the Philadelphia Neurodevelopmental Cohort (PNC) \cite{satterthwaite2014neuroimaging}. Here we study a sample of $n=879$ participants between the ages of 8 and 22 years (mean $=15.9$, s.d. $=3.3$, 386 males, 493 females) with high quality diffusion imaging, rest BOLD fMRI, and n-back task BOLD fMRI data. Our sample only contained subjects with low estimated head motion and without any radiological abnormalities or medical problems that might impact brain function (see Supplementary Information for detailed exclusion criteria). Details about imaging parameters, task design, and image preprocessing can be found in the Supplementary Information.
	
	\subsection*{Unsupervised clustering of BOLD volumes}
	
	BOLD fMRI activity patterns are known to represent information content \cite{Naselaris2011}, information processing \cite{Barch2013,Vossel2014}, and attention to stimuli \cite{Vossel2014}. Here, we use a discrete model as a simplification of brain dynamics \cite{Vidaurre2017}, in which we view repeatedly visited locations in regional activation space to be neural representations of cognitive states, or ``brain states'' for simplicity. In order to ultimately characterize the progression of these brain states from one time point to the next, and by extension the progression of the brain through regional activation space, we began by concatenating all functional volumes into one large data matrix \cite{Shine2019}. Specifically, we took all brain-wide patterns of BOLD activity from the resting-state scan and from the n-back task scan from all subjects, and we placed them into a matrix $\mathbf{X}$ with $N$ observations (rows) and $P$ features (columns). Here, $P$ is the number of brain regions in the parcellation (462), and $N$ is the number of subjects (879) $\times$ (120 resting state volumes $+$ 225 n-back task volumes), summing up to $N=303255$. 
	
	To determine the brain states present in these data, we performed 20 repetitions of $k$-means clustering for $k=2$ to $k=11$ using Pearson correlation as the algorithm's measure of distance \cite{Liu2013,Chen2015,Gutierrez-Barragan2019}. Because we aimed to study the temporal progression between coactivation patterns using a $k \times k$ transition probability matrix, and our resting state scans contained 120 frames, $k^2$ must be less than 120 in order to theoretically observe each transition at least once. Therefore, we chose $k=11$ ($k^2=121$) as our maximum possible value of $k$. After selecting $k=5$, we chose to consider the partition with the lowest error out of all 20 repetitions for subsequent analyses. \rev{To identify the optimal number of clusters $k$, we assessed the variance explained by the lowest error solution of the clustering algorithm at each value of $k$ from 2 to 11, and the gain in variance explained for a unit increase in $k$. The variance explained by the clustering algorithm is defined by the ratio of between-cluster variance to total variance in the data (within-cluster variance plus between-cluster variance) \cite{Gutierrez-Barragan2019,Goutte1999}. We also intended to make cross-subject comparisons of state dynamics as continuous measures, so it was important to use partitions that identified brain states that were common across all subjects, rather than identifying many different states that were each only represented in a few subjects.}
	
	\rev{We observed that the variance explained by the clustering algorithm began to taper off after $k=5$ (Fig. \ref{fig:figureS1}a), and the additional variance explained for each unit increase in $k$ after 5 was $<1\%$ (Fig. \ref{fig:figureS1}b). Additionally, $k$ values greater than $5$ produced states that were not all represented in every subject (Fig. \ref{fig:figureS1}c). To avoid using an unnecessarily large number of states while maintaining inter-subject correspondence in state presence, we chose $k=5$. To further validate the choice of $k=5$, we evaluated the split reliability of the partition at this resolution (Fig. \ref{fig:figureS1}d-f). This analysis showed that cluster centroids and transition probability matrices were highly similar between independently clustered subject samples (see Supplementary Methods for details). Another recent paper \cite{Gutierrez-Barragan2019} found a similar drop off in additional variance explained at $k=6$ instead of $k=5$. Key findings are reproduced at $k=6$ in the supplement and at $k=5$ for a second parcellation.}
	
	\subsection*{Analysis of spatiotemporal brain dynamics}
	
	After using $k$-means clustering to define discrete brain states, we generated names for each state using the maximum cosine similarity to binary vectors reflecting activation of communities in an \textit{a priori} defined 7-network partition \cite{ThomasYeo2011}; names were generated separately for maximum cosine similarity of positive and negative state entries. These names only serve as a convenient way of referring to clusters instead of their index (i.e., 1-$k$), and have no impact on any analyses. Next, we computed subject-level state \textit{fractional occupancy} as the percentage of volumes in each scan that were classified as a particular state. Additionally, we computed subject-level state \textit{dwell time} as the mean length of consecutive runs of each state. We defined the \textit{transition probability} between state $i$ and state $j$ to be the probability that $j$ is the next new state occupied after state $i$. This can also be equivalently framed as the probability of a specific state transition occurring given that some state transition is occurring. We chose this metric in order to understand state transitions without bias from potentially independent effects of state dwell time or autocorrelation. Operationally, this computation was performed by reducing the empirically obtained state sequences to a new sequence (e.g. [1 1 1 2 2 3 2 2] becomes [1 2 3 2]) in which the dwell time of every state is equal, and then computing the probability of state $j$ following state $i$. In the supplement, we also compute the \textit{transition probability} between two states as the probability of transitioning from state $i$ at time $t$ to state $j$ at time point $t+t_{r}$ given that the current state is $i$, where $t_r$ is the TR of the BOLD scanning sequence (3 seconds for PNC and 0.72 seconds for HCP) for the purposes of demonstrating the non-random nature of brain state dynamics. 
	
	Finally, in order to assess the context-dependent nature of brain state dynamics, we performed a non-parametric permutation test to compare group-average transition probabilities between the n-back task and the resting state. First, we randomly selected two halves of the full sample. Next, we generated two group-average transition matrices by averaging together resting state transition matrices from one half and n-back transition matrices from the other half, and \emph{vice versa}. This procedure was repeated 10000 times, and we retained the difference between the two halves at every element of the transition matrix. We generated a $p$-value for each element of the transition matrix by dividing the number of times the observed difference between n-back and rest at that element exceeded the null distribution of differences.
	
	\subsection*{Network control theory}
	
	To better understand the structural basis for the observed brain states themselves, as well as their persistence dynamics, we employed tools from network control theory \cite{gu2015controllability,Tang2017}.  We represent the fractional anisotropy-weighted structural network estimated from diffusion tractography as an $N\times N$ matrix $\mathbf{A}$, where $N$ is the number of brain regions in the parcellation and the elements $\mathbf{A}_{ij}$ contain the estimated strength of structural connectivity between region $i$ and $j$, where $i$ and $j$ can range from 1 to $N$. Because diffusion tractography cannot estimate within-region structural connectivity, $A_{ij} = 0$ whenever $i=j$.
	
	We allow each node to carry a real value, contained in the map $\mathbf{x} : \mathbb{R}_{\geq0} \rightarrow \mathbb{R}^{N}$, to describe the activity at each region in continuous time. Next, we employ a linear, time-invariant model of network dynamics:
	
	\begin{align}
		\dot{\mathbf{x}}(t) = \mathbf{Ax}(t) + \mathbf{B}\mathbf{u}(t)~,
	\end{align}\label{eq:cont}
	
	\noindent where $\mathbf{x}$ describes the activity (i.e. BOLD signal) in each brain region over time, and the value of the $i$\textsuperscript{th} element of $\mathbf{x}$ describes the activity level of region $i$.
	
	After stipulating this dynamical model, we computed the $k\times k$ transition energy matrix $\mathbf{T}_e$ as the minimum energy required to transition between all possible pairs of the $k$ clustered brain states, given the white matter connections represented in $\mathbf{A}$. See Supplementary Methods for details on computation of minimum control energy and selection of a control horizon. For the purposes of control theoretic simulations, we were interested in exploring the fundamental role of white matter architecture in supporting brain state transitions. Thus, we constructed a single group-representative $\mathbf{A}$ generated through distance-dependent consistency thresholding \cite{Betzel2017,Misic2015} of all subjects' structural connectivity matrices, a process which has been described in detail elsewhere \cite{Roberts2017a}.
	
	\subsection*{Developmental and cognitive trends of brain dynamics}
	
	After identifying non-random brain dynamics at the level of individual frames, we hypothesized that features of these dynamics would change throughout normative neurodevelopment, and moreover that they would map to cognitive performance. To assess potential developmental trends of spatiotemporal brain dynamics, we fit the following model using linear regression:
	
	\begin{equation} \label{eq:equation1}
	D = \beta_0 + \beta_{a} a + \beta_{v} v + \beta_{h} h + \beta_{m_{d}} m_{d} + \beta_{s} s + \epsilon~, 
	\end{equation} 
	
	\noindent where $a$ is age, $v$ is total intracranial volume, $m_{d}$ is the mean framewise displacement during rest or n-back scans, $h$ is handedness, $s$ is sex, $\epsilon$ is an error term, and $D$ is a measure of brain dynamics such as fractional occupancy, transition probability, or asymmetry. To assess potential relations between cognitive performance and spatiotemporal brain dynamics, we fit the following model using linear regression:
	
	\begin{equation} \label{eq:equation2}
	C =  \beta_0 + \beta_{D} D +  \beta_{a} a + \beta_{v} v + \beta_{h} h + \beta_{m_{d}} m_{d} + \beta_{s} s + \epsilon~, 
	\end{equation} 
	
	\noindent where $C$ is the overall or n-back block-specific $d^{\prime}$ score, which we use as our measure of working memory performance, and all other variables are the same as described above. For all analyses, we applied a Bonferroni correction for multiple comparisons, accounting for tests performed over all states or state transitions within each scan. We chose the Bonferroni-level correction because it is a conservative approach given that each state's fractional occupancies and transitions are not fully independent of one another.
	
	\section*{Data availability}
	
	All structural and functional neuroimaging data are available at \href{https://www.ncbi.nlm.nih.gov/projects/gap/cgi-bin/study.cgi?study_id=phs000607.v3.p2}{https://www.ncbi.nlm.nih.gov/projects/gap/cgi-bin/study.cgi?study\_id=phs000607.v3.p2}.
	
	\section*{Code availability}
	
	All analysis code is available at \href{https://github.com/ejcorn/brain_states}{https://github.com/ejcorn/brain\_states}.
	
	\section*{Acknowledgements}
	
	D.S.B. and E.J.C. acknowledge support from the John D. and Catherine T. MacArthur Foundation, the Alfred P. Sloan Foundation, the ISI Foundation, the Paul Allen Foundation, the Army Research Laboratory (W911NF-10-2-0022), the Army Research Office (Bassett-W911NF-14-1-0679, Grafton-W911NF-16-1-0474, DCIST- W911NF-17-2-0181), the Office of Naval Research, the National Institute of Mental Health (2-R01-DC-009209-11, R01 - MH112847, R01-MH107235, R21-M MH-106799), the National Institute of Child Health and Human Development (1R01HD086888-01), National Institute of Neurological Disorders and Stroke (R01 NS099348), and the National Science Foundation (BCS-1441502, BCS-1430087, NSF PHY-1554488 and BCS-1631550). T.D.S. acknowledges support from the National Institute of Mental Health (R01MH107703, R01MH113550, and RFMH116920). The content is solely the responsibility of the authors and does not necessarily represent the official views of any of the funding agencies.
	
	\newpage
	\bibliographystyle{brainstates2.bst}
	\bibliography{library.bib}
	\onecolumngrid
	\beginsupplement
	\newpage
	
	\section*{Supplementary information}
	\twocolumngrid
	\subsection*{Sample exclusion criteria}
	We excluded 722 of the initial 1601 subjects for the following reasons: medical problems that may impact brain function, incidental radiologic abnormalities in brain structure, poor or incomplete FreeSurfer reconstruction of T1 images \cite{Rosen2018}, high motion in rest or n-back fMRI scans, high signal-to-noise ratio or poor coverage in task-free or n-back task BOLD images, and failure to meet a rigorous manual and automated quality assurance protocol for DTI \cite{Roalf2016}. Notably, our goal in constructing a sample was to compare structure-function relationships between contexts across all subjects in our sample. This analysis required highly stringent inclusion criteria that only included subjects with high quality data for rest BOLD, n-back task BOLD, and DTI.
	
	\subsection*{Functional Scan Types}
	During the resting-state scan, a fixation cross was displayed as images were acquired. Subjects were instructed to stay awake, keep their eyes open, fixate on the displayed crosshair, and remain still. Total resting state scan duration was 6.2 minutes. As previously described \cite{Satterthwaite2012}, we used the fractal n-back task \cite{Ragland2002} to measure working memory function. The task was chosen because it is a reliable probe of the executive system and is not contaminated by lexical processing abilities that also evolve during adolescence \cite{Schlaggar2002,Brown2005}. The task involved the presentation of complex geometric figures (fractals) for 500 ms, followed by a fixed interstimulus interval of 2500 ms. This occurred under the following three conditions: 0-back, 1-back, and 2-back, inducing different levels of working memory load. In the 0-back condition, participants responded with a button press to a specified target fractal. For the 1-back condition, participants responded if the current fractal was identical to the previous one; in the 2-back condition, participants responded if the current fractal was identical to the item presented two trials previously. Each condition consisted of a 20-trial block (60 s); each level was repeated over three blocks. The target/foil ratio was 1:3 in all blocks, with 45 targets and 135 foils overall. Visual instructions (9 s) preceded each block, informing the participant of the upcoming condition. The task included a total of 72 s of rest, while a fixation crosshair was displayed, which was distributed equally in three blocks of 24 s at the beginning, middle, and end of the task. Total task duration was 693 s. To assess performance on the task, we used $d^\prime$, a composite measure that takes into account both correct responses and false positives to separate performance from response bias \cite{snodgrass1988pragmatics}.
	
	\subsection*{Imaging data acquisition and preprocessing}
	
	MRI data were acquired on a 3 Tesla Siemens Tim Trio whole-body scanner and 32-channel head coil at the Hospital of the University of Pennsylvania. High-resolution T1-weighted images were acquired for each subject. For diffusion tensor imaging (DTI), 64 independent diffusion-weighted directions with a total of 7 $b=0$ s/mm\textsuperscript{2} acquisitions were obtained over two scanning sessions to enhance reliability in structural connectivity estimates \cite{Satterthwaite2014}. All subjects underwent functional imaging (TR $= 3000$ ms; TE $= 32$ ms; flip angle $= 90$ degrees; FOV = $192 \times 192$ mm; matrix = $64 \times 64$; slices $= 46$; slice thickness $= 3$ mm; slice gap $= 0$ mm; effective voxel resolution = $3.0 \times 3.0 \times 3.0$ mm) during the resting-state sequence and the n-back task sequence \cite{Satterthwaite2014,Satterthwaite2013}. During resting-state and n-back task imaging sequences, subjects' heads were stabilized in the head coil using one foam pad over each ear and a third pad over the top of the head in order to minimize motion. Prior to any image acquisition, subjects were acclimated to the MRI environment via a mock scanning session in a decommissioned scanner. Mock scanning was accompanied by acoustic recordings of gradient coil noise produced by each scanning pulse sequence. During these sessions, feedback regarding head motion was provided using the MoTrack motion tracking system (Psychology Software Tools, Inc., Sharpsburg, PA). 
	
	Raw resting-state and n-back task fMRI BOLD data were preprocessed following the most stringent of current standards \cite{Satterthwaite2013a,Ciric2017} using XCP engine \cite{Ciric2018}: (1) distortion correction using FSL's FUGUE utility, (2) removal of the first 4 volumes of each acquisition, (3) template registration using MCFLIRT \cite{Jenkinson2002}, (4) de-spiking using AFNI's 3DDESPIKE utility, (5) demeaning to remove linear or quadratic trends, (6) boundary-based registration to the individual high-resolution structural image, (7) 36-parameter global confound regression including framewise motion estimates and signal from white matter and cerebrospinal fluid, and (8) first-order Butterworth filtering to retain signal in the 0.01 to 0.08 Hz range. Following these preprocessing steps, we parcellated the voxel-level data using the 463-node Lausanne atlas \cite{Cammoun2012}. We excluded the brainstem, leaving 462 parcels. Our choice of parcellation scale was motivated by prior work showing that parcellations of this scale replicate voxelwise clustering results more than coarser scales with fewer parcels \cite{Chen2015}. We excluded any subject with mean relative framewise displacement $> 0.5$ mm or maximum displacement $> 6$ mm during the n-back scan, and mean relative framewise displacement $> 0.2$ mm for the resting state scan.
	
	All DTI datasets were subject to a rigorous manual quality assessment protocol that has been previously described \cite{Roalf2016}. The skull was removed by applying a mask registered to a standard fractional anisotropy map (FMRIB58) to each subject's DTI image using an affine transformation. The FSL EDDY tool was used to correct for eddy currents and subject motion and rotate diffusion gradient vectors accordingly. Distortion correction was applied using FSL's FUGUE utility. DSI studio was then used to estimate the diffusion tensor and perform deterministic whole-brain fiber tracking with a modified FACT algorithm that used exactly 1,000,000 streamlines per subject excluding streamlines with length $< 10$ mm \cite{Baum2017,Gu2015}. Lausanne 463-node atlas parcels were extended into white matter with a 4 mm dilation \cite{Baum2017,Gu2015} and then registered to the first $b = 0$ volume using an affine transform \cite{Roalf2016,Baum2017}. For all analyses, edge weights in the structural network were defined by the average fractional anisotropy value for streamlines connecting each pair of parcels \cite{Baum2017a}.
	
	\subsection*{Split-halves validation of clustering}
	
\begin{figure*}
		\begin{center}
			\includegraphics[width=18cm,keepaspectratio]{\string 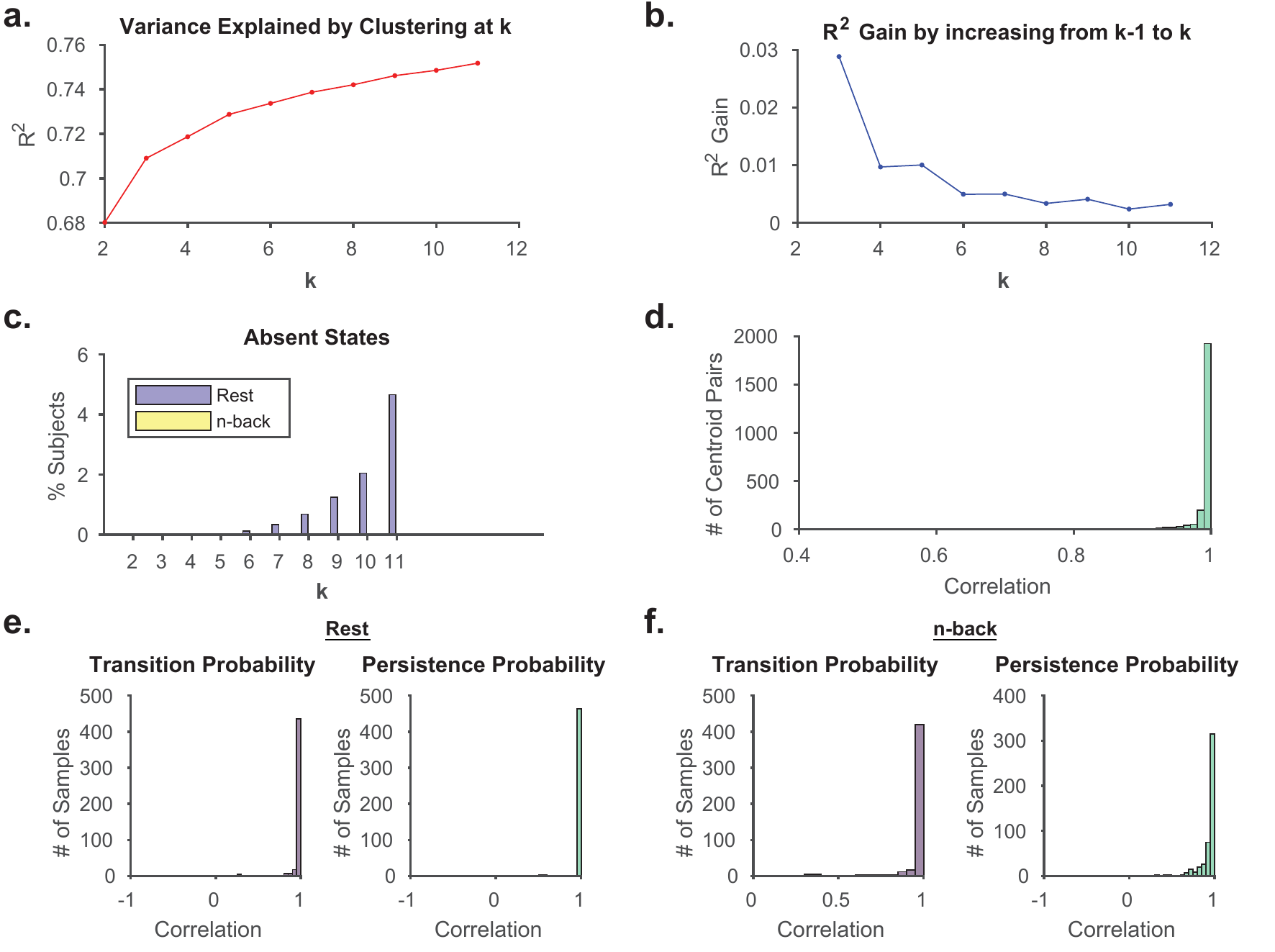}
			\caption{\textbf{Choosing the number of clusters.} \emph{(a)} Mean (line) and standard deviation (error bars) of $z$-scored rand index (ZRI) values calculated for every pair of cluster partitions in 100 repetitions of $k$-means on concatenated rest and n-back BOLD data for $k=2$ to $k=11$. \emph{(b)} Percentage of subjects missing at least one state for rest and n-back. At $k>5$, states begin to be incompletely represented across subjects. \emph{(c-e)} Split-halves validation of cluster centroids \emph{(c)}, and state transition and persistence probabilities for rest and n-back \emph{(d-e)}. Pearson correlation coefficients between transition and persistence probabilities for each pair of split cluster partitions are predominantly $r>0.99$, thus showing a high degree of robustness to sample composition in estimating brain states and their dynamics. \emph{ZRI}, $z$-scored rand index.
				\label{fig:figureS1}}
		\end{center}
	\end{figure*}

	To ensure that our final clustering solution for $k=5$ was not influenced by outliers or adversely impacted by model overfitting, we split our sample into two equal partitions 500 times and performed $k$-means clustering separately on each half of the dataset. We then matched clusters by computing the cross-correlation between both sets of centroids, and then by reordering the clusters based on the maximum correlation value for each cluster. We plotted those maximum correlation values and found that most were $>0.99$, suggesting a high degree of robustness and stability in brain states (Fig. \ref{fig:figureS1}d). We also computed the state transition probabilities and state persistence probabilities for each half separately for rest and n-back, and then computed the correlation between the transition or persistence probabilities between the two data set partitions. Similarly, we found very high correlation values ($> 0.99$) for state transition probabilities and state persistence probabilities for both rest and n-back (Fig \ref{fig:figureS1}e-f). These observations suggest that our estimates of brain dynamics are robust to outliers and consistent across different subsamples of our data.
	
	\subsection*{Functional connectivity does not fully explain instantaneous coactivation}
	
	We posit that the alignment between \textit{a priori} resting state functional networks \cite{ThomasYeo2011} (RSNs) and our coactivation patterns cannot be fully explained by analyzing interregional correlations in BOLD signal across time, which is often called ``functional connectivity.'' Here, we provide evidence that suggests that our coactivation patterns are consistent with but not trivially explained by functional connectivity. We also provide examples that explain why temporal correlation does not necessarily explain instantaneous coactivation in general, and demonstrate that $k$-means clustering is a useful tool for extracting coactivation patterns.
	
\begin{figure*}
		\begin{center}
			\includegraphics[width=18cm,keepaspectratio]{\string 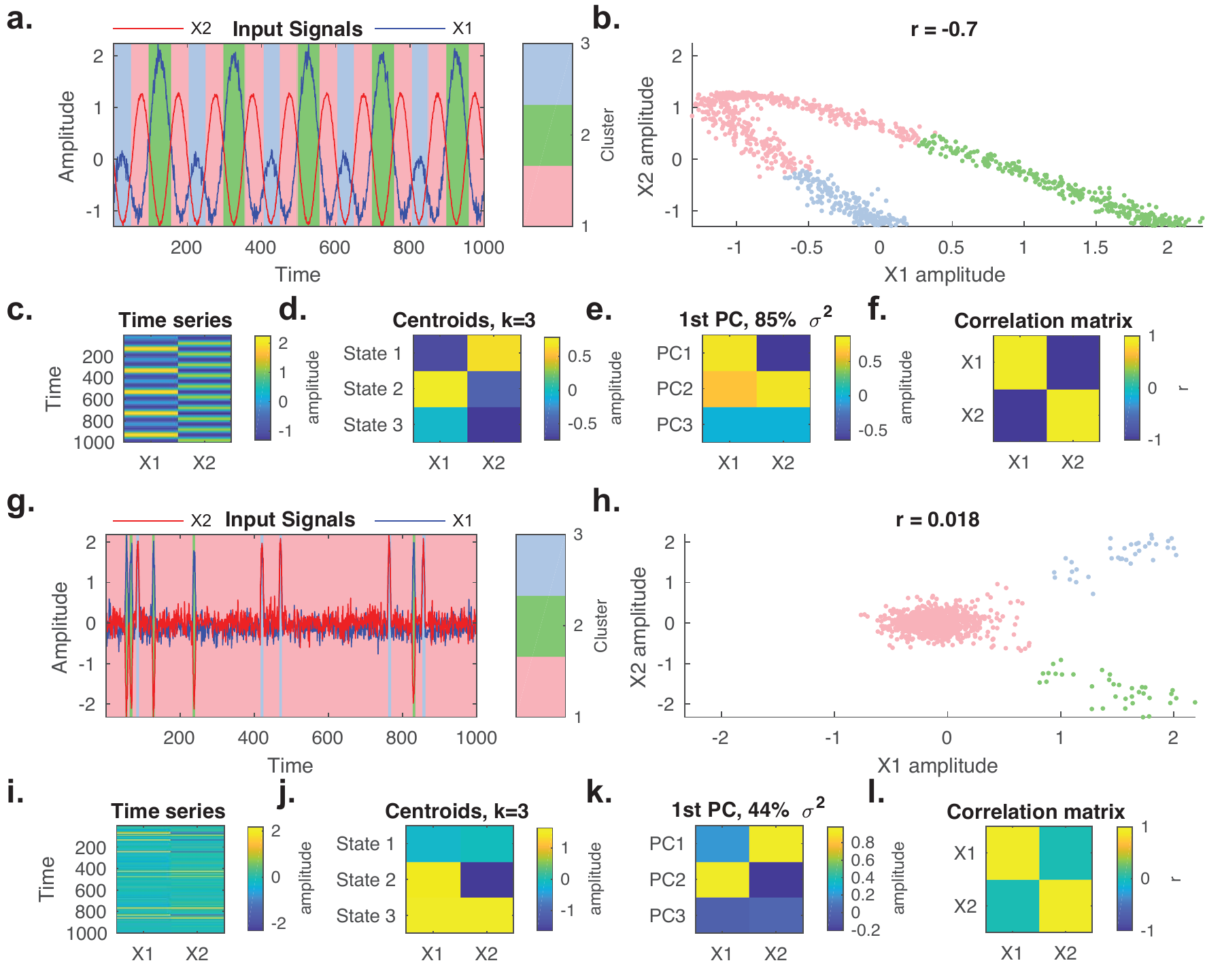} 
			\caption{\textbf{Instantaneous coactivation can deviate from temporal correlation patterns.} \emph{(a)} Signal vector X1, generated by $sin(t)$ for $t=1:1000$, plus Gaussian noise is plotted in blue. Signal vector X2, generated by subtracting X1 from a slower sin wave, plus Gaussian noise, is plotted in red. One can observe three major coactivation states, which can be extracted through $k$-means clustering (partition shown in light red, green, and blue). \emph{(b)} Plot of X1 amplitude ($x$-axis) vs. X2 amplitude ($y$-axis), yielding a Pearson's $r$ of -0.70. \emph{(c)} A $T \times N$ data matrix $\mathbf{X_a}$ consisting of length $T=1000$ time series for $N=2$ channels, constructed by concatenating X1 and X2. We concatenated $\mathbf{X_a}$ with two channels of Gaussian noise to generate a $T\times4$ data matrix $\mathbf{X_n}$. The additional noise channels allow correlation to be used as the distance metric, as in the submitted manuscript. \emph{(d)} Centroids, or ``states,'' obtained through $k$-means clustering of $\mathbf{X_n}$, using a distance metric of correlation as in the main text. Three states emerge; the first state consists of low- and high-amplitude activity in X1 and X2, respectively. The second state consists of low amplitude activity in X2 with near 0 activity in X1. The third state consists of high amplitude activity in X1 with low amplitude activity in X2. \emph{(e)} The first 3 principal component (PC) loadings of $\mathbf{X_a}$. PC1 captures States 1 and 2, but no single PC captures State 3. \emph{(f)} The cross-correlation structure of $\mathbf{X_a}$, which does not trivially reveal the states in panel \emph{(d)}. \emph{(g)} Signal vector X1, generated by Gaussian noise plus random positive sinusoidal activations is plotted in blue. Signal vector X2, generated by Gaussian noise plus synchronized coactivation or coinactivation with X1, is plotted in red. One can observe three major coactivation states, which can be extracted through $k$-means clustering (partition shown in light red, green, and blue). \emph{(h)} Plot of X1 amplitude ($x$-axis) vs. X2 amplitude ($y$-axis), yielding a Pearson's $r$ of 0.018. \emph{(i)} A $T \times N$ data matrix $\mathbf{X_b}$ consisting of length $T=1000$ time series for $N=2$ channels, constructed by concatenating X1 and X2. We concatenated $\mathbf{X_b}$ with two channels of Gaussian noise to generate a $T\times4$ data matrix $\mathbf{X_{m}}$. The additional noise channels allow correlation to be used as the distance metric, as in the submitted manuscript. \emph{(j)} Centroids, or ``states,'' obtained through $k$-means clustering of $\mathbf{X_{m}}$, using a distance metric of correlation as in the main text. Three states emerge; the first consists of high-amplitude activity in X1 with low-amplitude activity in X2. The second consists of near 0 activity in X1 and X2. The third consists of high-amplitude activity in both X1 and X2. \emph{(k)} The first 3 principal component (PC) loadings of $\mathbf{X_b}$. PC1 does not exist within the data; PC2 mirrors State 1; and both States 2 and 3 are not reflected in any PCs. \emph{(l)} Cross-correlation structure of $\mathbf{X_b}$, which does not trivially reveal the cluster centroids in panel \emph{(d)}.
		\label{fig:figureS11}}
		\end{center}
	\end{figure*}

	In general, time series data can contain (1) unexpected coactivation patterns in the presence of temporal correlation and (2) unexpected coactivation patterns in the absence of temporal correlation. Here, we provide two examples that illustrate how $k$-means clustering can resolve dissociation between temporal correlation and instantaneous coactivation in multi-dimensional time series data. First, we show two stationary signals that are anticorrelated across time (Fig. \ref{fig:figureS11}b, $r=-0.70$), but exhibit three unexpected coactivation patterns (rather than two simple ``on-off'' and ``off-on'' patterns) due to fluctuation between variable amplitudes (Fig. \ref{fig:figureS11}a-b). In this example, one sinusoidal signal fluctuates between three distinct amplitudes and the other sinusoidal signal simply fluctuates between a peak and a trough (Fig. \ref{fig:figureS11}a). The combination of these signals yields 3 distinct coactivation patterns, rather than the 2 patterns that would be expected from a pair of temporally anticorrelated signals (Fig. \ref{fig:figureS11}c-e).
	
	Next, we show two signals that are uncorrelated across time, but still have multiple spontaneous coactivation patterns (Fig. \ref{fig:figureS11}a). Because these two signals sometimes activate together and at other times exhibit opposing activity, their temporal correlation is near 0 ($r=0.018$), yet there are 3 unique coactivation patterns in the time series that $k$-means faithfully extracts. Positive temporal correlations could occur either through simultaneous activation or simultaneous inactivation, while temporal anticorrelation can result from opposing activity patterns. One could envision how adding in additional signals (i.e. dimensions or brain regions) could provide additional degrees of freedom to construct a range of coactivation patterns for a given temporal correlation structure. The sensitivity of the $k$-means approach to coactivation patterns in these two contexts supports its use in fMRI data for identifying previously unknown spontaneous interactions between neural networks, as well as the temporal organization of those networks.
	
\begin{figure*}
		\begin{center}
			\includegraphics[width=17cm,keepaspectratio]{\string 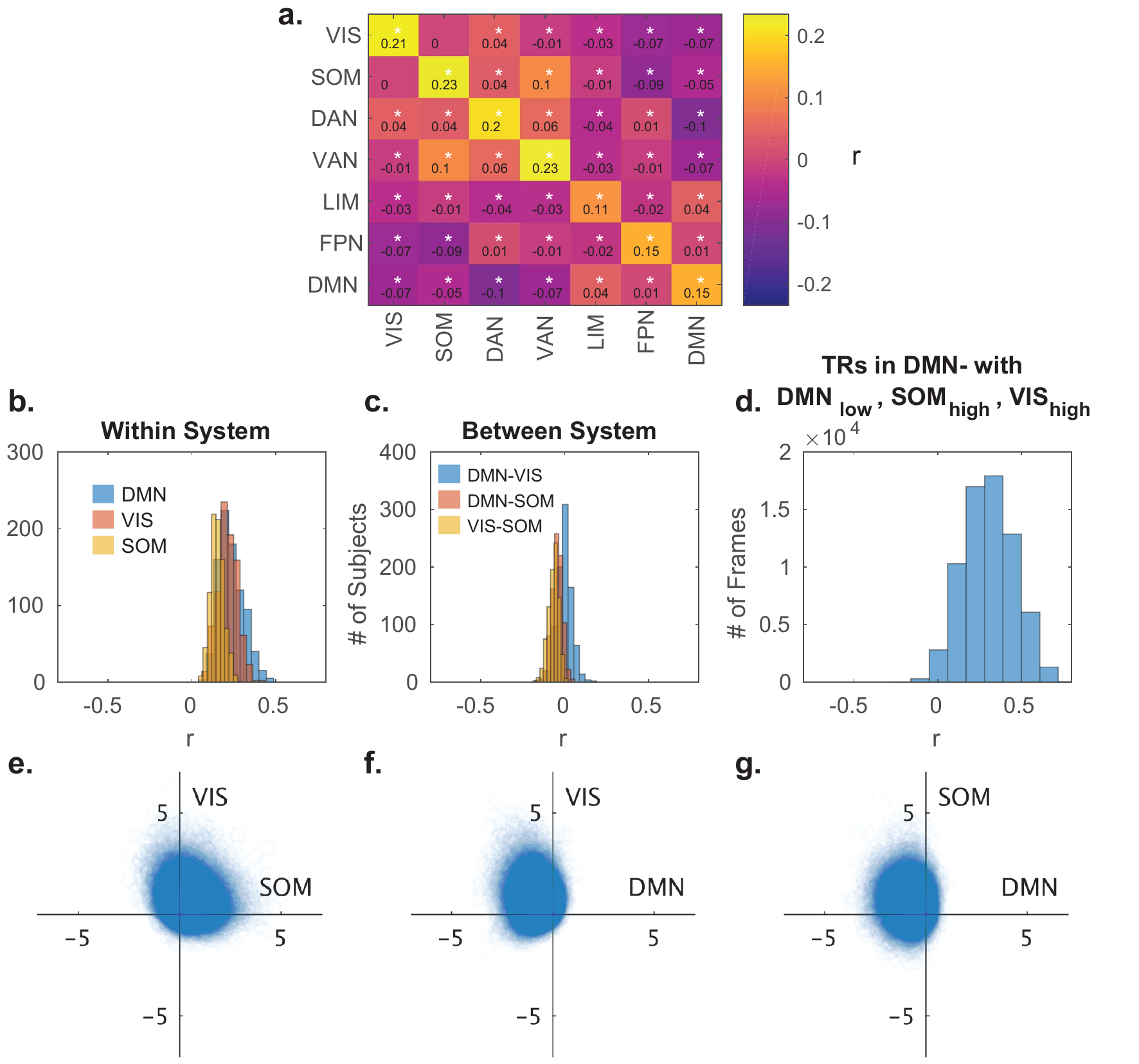}
			\caption{\textbf{Functional connectivity does not fully explain spontaneous coactivation.} \emph{(a)} Group average, mean functional connectivity within and between cognitive systems defined \textit{a priori} \cite{ThomasYeo2011}. We first computed average functional connectivity matrices by averaging functional connectivity matrices over rest and n-back scans for each subject. Next, we computed the average edge strength within and between regions in each \textit{a priori} defined system to yield mean within- and between-system functional connectivity. Finally, we computed the group average mean within- and between-system functional connectivity across the $n=879$ subjects from the main text, the values of which are displayed in black text overlaying each heatmap element. *, $p<0.05$ for a one-sample $t$-test comparing the mean of the distribution of correlations across subjects to 0. \emph{(b)} Mean temporal correlation values within regions belonging to the default mode network (DMN), visual system (VIS), and somatomotor system (SOM) show positive correlations within systems on average. \emph{(c)} Mean temporal correlations between DMN, VIS, and SOM regions show weakly negative (DMN-VIS, DMN-SOM) or near 0 (VIS-SOM) correlations on average. \emph{(d)} Histogram of TRs (individual BOLD frames) exhibiting activity patterns with low activity in the DMN and high activity in the SOM and VIS systems, as measured by correlation with a binary indicator vector. Despite low temporal correlations, we still find activity patterns with spatial anticorrelation between these systems. \emph{(e-g)} BOLD data in regional activation space. Scatter plots where points are individual BOLD TRs in the DMN- cluster and axes reflect the mean activity across regions in the DMN, VIS, or SOM for each TR, shown in 2 dimensions for ease of visualization. These plots show that TRs in the DMN- cluster (Fig. 2a) have simultaneous high activity in VIS and SOM (panel \emph{d}) in addition to low activity in the DMN (panel \emph{e-f}).
					\label{fig:figureS12}}
		\end{center}
	\end{figure*}

	When we apply $k$-means clustering to high-dimensional BOLD data from resting state and n-back scans (Fig. \ref{fig:figure2}a), we identify activity patterns with both expected and unexpected features based on functional connectivity. In general, we saw that RSNs show coherent high or low amplitude activity in each state (Fig. \ref{fig:figureS12}c). This finding reflects the strong positive correlations \textit{within} RSNs (Fig. \ref{fig:figureS12}a). We also see coactivation patterns consistent with patterns of functional connectivity \textit{between} RSNs. For instance, the DMN+ state (Fig. \ref{fig:figure2}a) shows spatial anticorrelation between the temporally anticorrelated dorsal attention network and default mode network \cite{Fox2005} (Fig. \ref{fig:figureS12}a, mean $r=-0.10$, one-sample $t$-test, $df=878$, $t=-80.45$, $p<10^{-15}$). However, we also identify coactivation patterns that do not trivially reflect functional connectivity. The DMN- state centroid (Fig. \ref{fig:figure2}a) consists of low amplitude activity throughout the default mode network (DMN) with high amplitude activity throughout somatomotor and visual systems (Fig. \ref{fig:figure2}c). When clustering on functional connectivity \cite{ThomasYeo2011}, these three systems emerge as separate. Mean correlations within these systems are positive (Fig. \ref{fig:figureS12}a-b), but correlations between them are near zero (SOM and VIS) or weakly negative (DMN with SOM and VIS) (Fig. \ref{fig:figureS12}c). Based solely on functional connectivity, one would not expect these three systems to strongly coactivate or oppose one another at the level of individual BOLD frames (TRs), yet our analysis suggests that there are many TRs with low DMN activity, high VIS activity, and high SOM activity (Fig. \ref{fig:figureS12}d-g). Moreover, VIS and SOM have two additional configurations of coactivation represented in the VIS+ and VIS- states (Fig. \ref{fig:figure2}a-c). In the VIS- state, we see high amplitude SOM activity with low amplitude VIS activity, suggesting that at times VIS and SOM systems coactivate and at other times they oppose one another. In the VIS+ state we see low amplitude SOM activity with high amplitude VIS activity. Purely based on functional connectivity, one might inappropriately draw the conclusion that these two RSNs are independent given their mean correlation of zero, when in reality they have 3 distinct configurations at the level of individual time points. The behavior of VIS and SOM is most consistent with the scenario presented in Fig. \ref{fig:figureS11}. Overall, these findings suggest that our analysis identifies recurrent activity patterns whose spatial organization reflects strong temporal correlations within RSNs, but also with coactivation between RSNs that cannot be trivially explained by temporal correlations between RSNs.
	
	\subsection*{Silhouette analysis of clustering}
	
\begin{figure*}
		\begin{center}
			\includegraphics[width=18cm,keepaspectratio]{\string 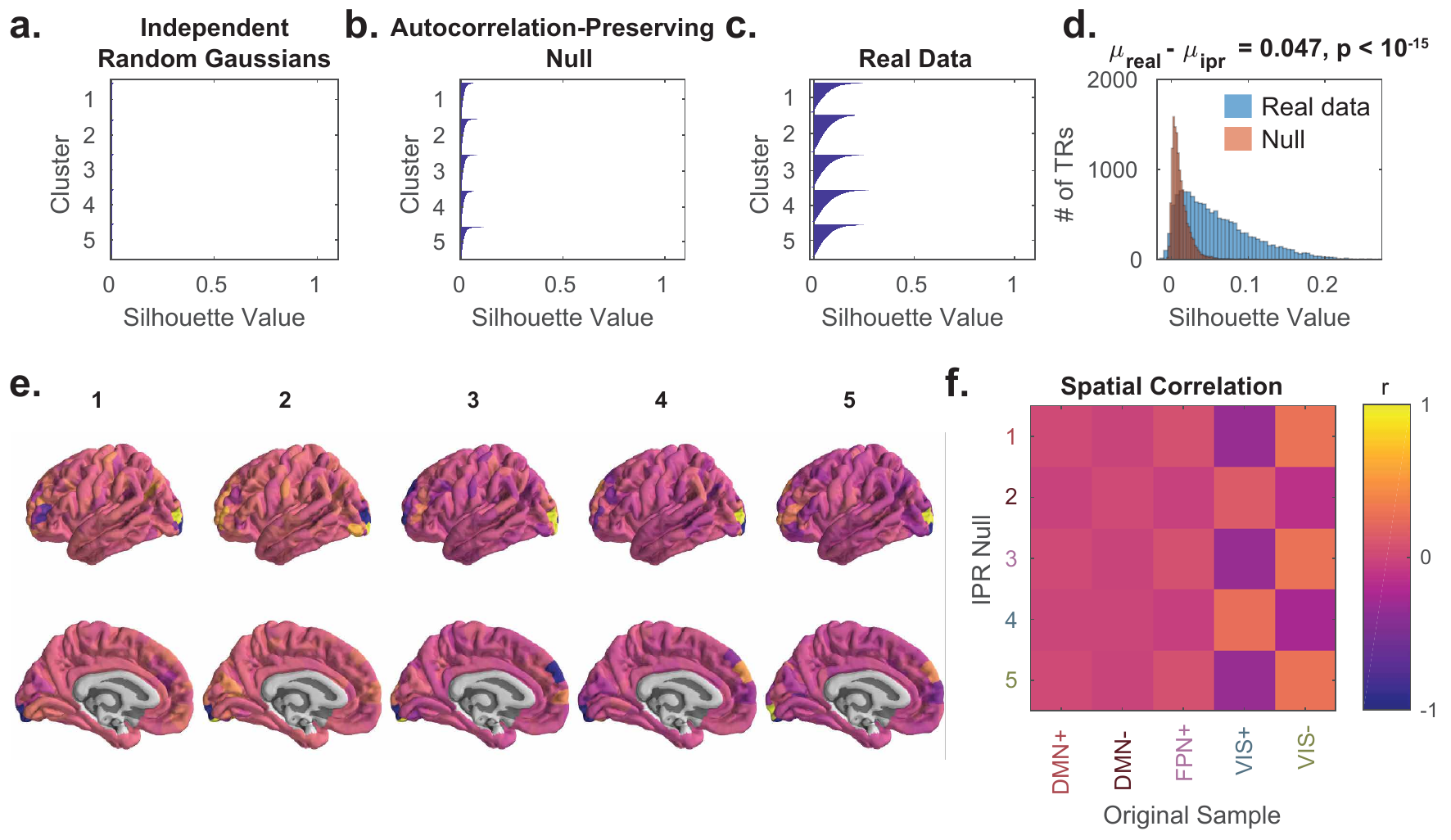}
			\caption{\textbf{BOLD data exhibits clustering in regional activation space.} We performed $k$-means clustering on rest and n-back data from the $n=879$ sample studied in the main text at $k$=5 using the correlation distance in a 462-region activation space. Here we show silhouette scores for rest and n-back time points from 40 random subjects; however, results are consistent across random sets of 10 subjects, suggesting marked reliability. Silhouette scores range from -1 to 1 and are computed for each data point, with 1 indicating that a data point is closer to members of its assigned cluster than to members of the next closest cluster, 0 indicating equidistance between the assigned cluster and the closest cluster, and -1 indicating that the data point is closer to another cluster. \emph{(a)} Silhouette values for data points from 462 independent, normally distributed channels. \emph{(b)} Silhouette values for data from an independent phase randomized (IPR) null model applied separately to resting state and n-back BOLD data. This null model preserves regional autocorrelation while eliminating non-stationarities and reducing covariance. \emph{(c)} Silhouette values for actual resting state and n-back BOLD data. \emph{(d)} Distribution of silhouette values across all clusters for a null model preserving autocorrelation (red, $\mu_{ipr}$) and for actual data (blue, $\mu_{real}$). A two-sample $t$-test confirms that silhouette values are larger in the real data, indicating that the data are clustered in regional activation space beyond what is expected from a signal with the same regional autocorrelations. \emph{(e)} Cluster centroids at $k=5$ using IPR resting state and n-back BOLD data. \emph{(f)} Spatial correlation between IPR null centroids in panel \emph{e} ($y$-axis) and full sample centroids ($x$-axis).
					\label{fig:figureS13}}
		\end{center}
	\end{figure*}

	In order to support the use of a discrete model of brain dynamics, we asked whether individual BOLD time points from rest and n-back scans exhibited clustering in a 462-region activation space. We used the silhouette scores of BOLD time points as a measure of clustering in this space. Silhouette scores range from -1 to 1 and are computed for each data point, with 1 indicating that a data point is closer to members of its assigned cluster than to members of the next closest cluster, 0 indicating equidistance between the assigned cluster and the closest cluster, and -1 indicating that the data point is closer to another cluster. We compared the silhouette scores for real BOLD data points from the PNC, data points from 462 independent random Gaussian distributions, and data points from independent phase randomized null time series \cite{Liegeois2017} based on subject-specific BOLD data. This independent phase randomized null model (IPR) preserves the autocorrelation within each region, but destroys covariance between regions. When we compared the silhouette scores for clustering of real BOLD data to independent random Gaussian and autocorrelation-preserving null data, we found that the real data had higher mean silhouette scores than that of the autocorrelation-preserving null data (Fig. \ref{fig:figureS13}a, difference in mean silhouette score actual minus null $=0.047$, two sample $t$-test, $df = 27598 $, $t=115.03$, $p < 10^{-15}$). Additionally, the centroids generated by clustering this null data had little obvious structure (Fig. \ref{fig:figureS13}e) and showed little similarity to the original centroids (Fig. \ref{fig:figureS13}f). These findings suggest that BOLD data exhibit non-trivial clustering in regional activation space.
	
	\subsection*{Impact of scan composition on brain states and dynamics}
	
\begin{figure*}
		\begin{center}
			\includegraphics[width=18cm,keepaspectratio]{\string 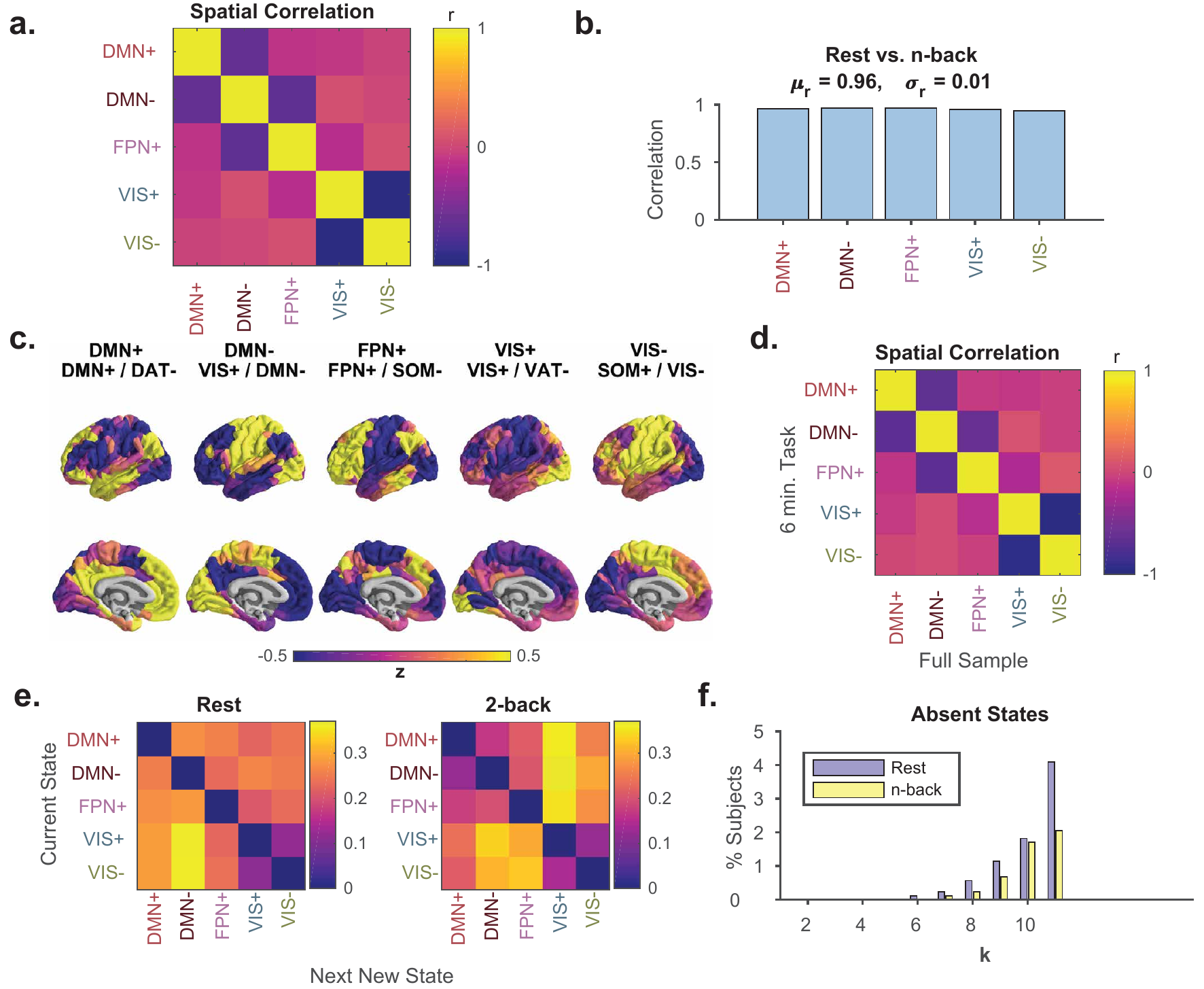}
			\caption{\textbf{Similarity between states in rest and n-back.} \emph{(a)} Spatial correlation between cluster centroids reveals anticorrelation between DMN- and DMN+, between DMN- and FPN+, and between VIS+ and VIS-. \emph{(b)} Spatial correlation between centroids calculated separately for rest and n-back reveal high correspondence, consistent with the identification of recurrent activity patterns common to both scans. \emph{(c)} Cluster centroids computed by including equal amounts of rest and n-back task data as input to the clustering algorithm. Cluster names based on maximum cosine similarity were identical to the full sample centroids. \emph{(d)} Spatial correlation between the 6 minute rest and the 6 minute n-back task cluster centroids and the full sample cluster centroids. Correlation coefficients of $r>0.99$ were found only on the diagonal, suggesting 1-to-1 correspondence between the two centroid sets. The observed off-diagonal anticorrelations are consist with those observed in the full sample, as shown in panel \emph{(a)}. \emph{(e)} Group average state transition probabilities for rest (\emph{right}) and n-back (\emph{left}) using the 6 minute n-back task cluster partition reveals similar structure and high correlation with full sample state transition probabilities. \emph{(f)} The $y$-axis shows the percentage of subjects missing at least one state in their time series for rest (\emph{purple}) and for n-back (\emph{yellow}), for values of $k$ on the $x$-axis ranging from 2 to 18. There existed at least one subject with missing states for all $k>5$, supporting the choice of $k=5$ for the main text.
				\label{fig:figureS2}}
		\end{center}
	\end{figure*}

	To ensure that our results were not biased by the fact that there were a larger number of n-back volumes (225 per scan) than rest volumes (120 per scan), we used the partition generated by clustering both entire scans together to compute separate centroids for volumes in rest or n-back scans. This analysis revealed a mean spatial Pearson correlation of $0.96$ between corresponding centroids (Fig. \ref{fig:figureS2}b). Next, we generated a new sample by concatenating the first 6 minutes of the n-back task data for each subject and the entire 6 minutes of the rest data for each subject. We ran this sample through the clustering algorithm at $k=5$ and found that the cluster centroids (Fig. \ref{fig:figureS2}c) were highly similar to those computed from the full sample (mean Pearson $r = 0.99$; Fig. \ref{fig:figureS2}d). We also computed transition probabilities using this cluster partition and identified highly similar group average transition matrix structure (rest, Pearson $r = 0.997$, n-back, Pearson $r = 0.989$, Fig. \ref{fig:figureS2}e), suggesting that the temporal order of state labels was largely unaffected by the scan composition of the sample. Moreover, these results suggest that n-back state transitions are internally consistent. 
	
	Finally, we show that the differences between rest and n-back in the proportion of subjects with any absent states (Fig. \ref{fig:figureS2}f) is attenuated relative to the full sample (Fig. \ref{fig:figureS1}e). This finding suggests that in the full sample, the n-back task data has better state representation due to better sampling, rather than poor classification of rest volumes. However, even with equal samples, the rest dataset still has more subjects with missing states, suggesting that there may be more variability in brain dynamics during rest. Importantly, there were still no subjects with absent states for rest or n-back at $k<5$ (Fig. \ref{fig:figureS2}f), and there was at least 1 subject with a missing state in rest or n-back at $k>5$ (though bars are very small in Fig. \ref{fig:figureS2}f). Collectively, these results support the simultaneous generation of partitions for rest and n-back volumes and the choice of $k=5$ for analysis in the main text.
	
	\subsection*{Impact of sampling rate, global signal regression, and head motion on brain states and dynamics}
	
	The BOLD data from the PNC was acquired at a sampling rate of one volume every 3 seconds \cite{satterthwaite2014neuroimaging}, which is relatively slow compared with other large data sets, including the Human Connectome Project \cite{VanEssen2013}, which samples every 0.72 seconds. The standard preprocessing pipeline for this data set involves regression of head motion parameters, white matter confounds, cerebrospinal fluid confounds, and global signal from each voxel's time series \cite{Ciric2018,Xia2017}. It is controversial whether this procedure, known as ``global signal regression,'' induces anticorrelation \cite{Chai2012,Murphy2017}.
	
	Thus, we selected 100 unrelated subjects from the minimally preprocessed version of the Human Connectome Project (HCP) data set \cite{Glasser2013} and performed the following preprocessing steps on resting state and n-back working memory task scans: (1) head motion regression, (2) linear and quadratic detrending, (3) bandpass filtering to retain the 0.01 to 0.08 Hz range, and (4) parcellation according to the 462 region Lausanne atlas. We concatenated all 405 volumes from the working memory task with the first 405 volumes from the resting state over all 100 subjects. We chose to make the number of volumes from each scan equal so that the clustering algorithm would not be biased towards one scan or the other. 
	
\begin{figure*}
		\begin{center}
			\includegraphics[width=18cm,keepaspectratio]{\string 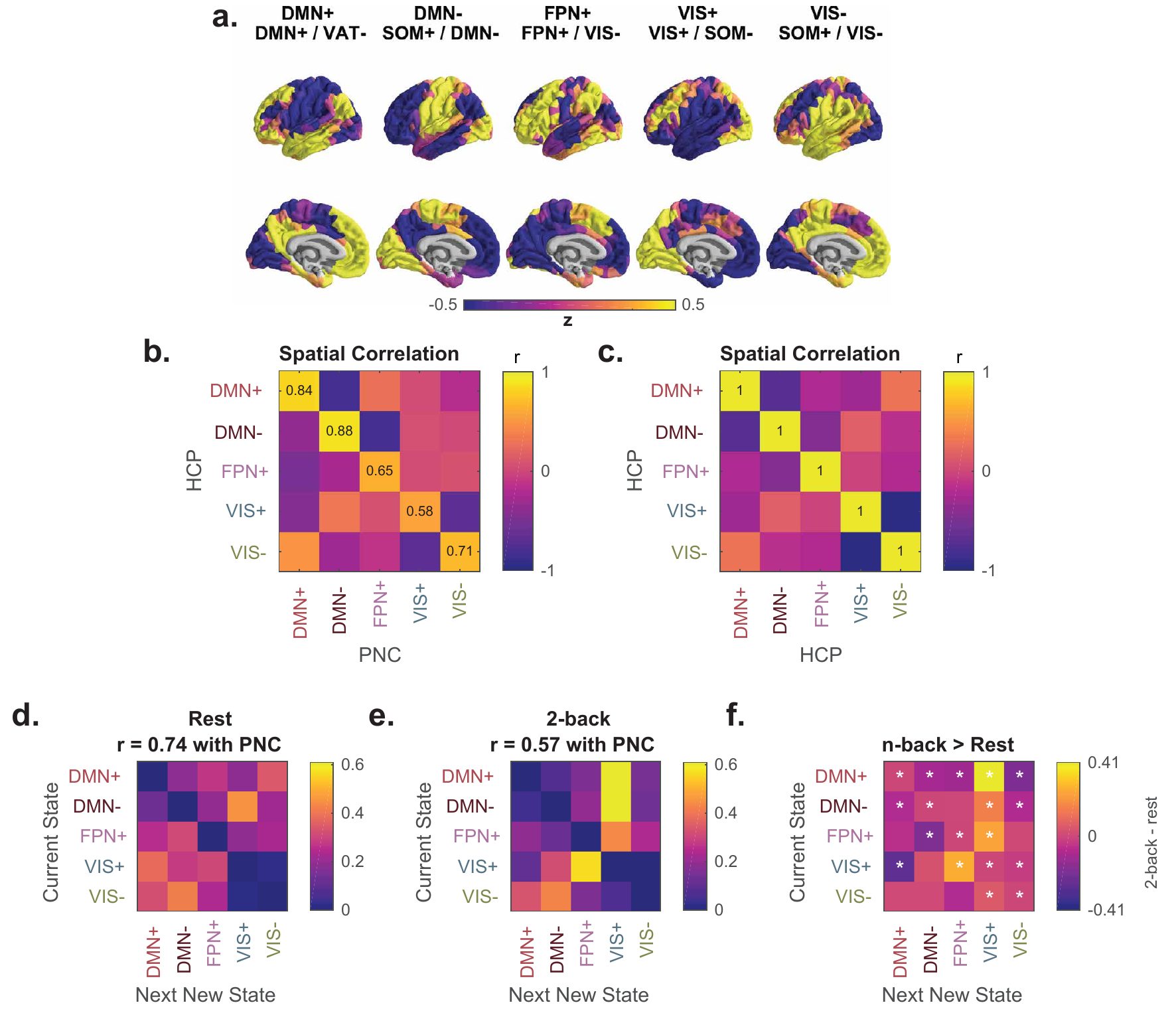}
			\caption{\textbf{Brain states and dynamics in an independent sample with higher sampling rate and no global signal regression.} \emph{(a)} Cluster centroids for clustering of rest and n-back task BOLD data from the Human Connectome Project (HCP) with volumes acquired 4 times as frequently as the PNC and no global signal regression. \emph{(b)} Spatial correlation between centroids for HCP and PNC data sets, is high along the diagonal, allowing for unambiguous matching of brain states between the two samples. \emph{(c)} Spatial correlation between HCP centroids. DMN+ and DMN-, along with VIS+ and VIS-, exhibit strong anticorrelation. \emph{(d-e)} HCP group average state transition probability matrices for rest \emph{(d)} and n-back \emph{(e)} scans. Off-diagonal elements of HCP rest and n-back transition matrices exhibit Pearson correlations of $r=0.83$ and $r=0.76$ with the PNC, respectively. HCP persistence probabilities are correlated with PNC persistence probabilities at $r=0.85$ and $r=0.85$ for rest and n-back, respectively. \emph{(f)} Non-parametric permutation testing demonstrating differences between the rest and n-back group average transition probabilities and persistence probabilities. Extremes of the color axis indicate statistical significance, with larger values indicating higher transition probabilities in n-back relative to rest. *, Bonferroni-adjusted $p < 0.05$ or $p > 0.95$.
				\label{fig:figureS3}}
		\end{center}
	\end{figure*}

	Next, we performed $k$-means clustering on this matrix and computed the centroids (Fig. \ref{fig:figureS3}a). Every HCP centroid was maximally correlated with only one PNC centroid, and \emph{vice versa}, allowing for unambiguous matching between the two sets of brain states (Fig. \ref{fig:figureS3}b). The DMN+ and DMN- states were the most similar to PNC states, while VIS+ and VIS- exhibited slightly lower correlations (Fig. \ref{fig:figureS3}b). DMN+ and DMN- states, as well as VIS+ and VIS- states, exhibited strong anticorrelation with each other (Fig. \ref{fig:figureS3}c). Nevertheless, the HCP off-diagonal elements of the transition probability matrices (i.e. transitions not persistence) for rest and 2-back block of the n-back task were correlated with PNC transition probabilities at $r=0.74$ and $r=0.57$, respectively (Fig. \ref{fig:figureS3}d-e). This finding suggests that while there were differences in the spatial activity patterns of brain states, their dynamic progression through time was relatively similar. The unexplained variance between the two samples could be due to differences in age, with the PNC comprised of developing youths and the HCP comprised of healthy adults. Notably, task dynamics were less similar between the two groups, possibly reflecting stronger age-related changes in task dynamics relative to resting state. Moreover, the differences in transition probabilities between 2-back and rest were highly similar in HCP and in PNC (Fig. \ref{fig:figureS3}e), with increased transitions from DMN and FPN states into VIS states. Overall, these findings suggest that while global signal regression and sampling rate may impact the spatial activity patterns comprising brain states to some degree, it does not impact estimation of their dynamics or the presence of spatial anticorrelation in their activity patterns. 
	
\begin{figure*}
		\begin{center}
			\includegraphics[width=18cm,keepaspectratio]{\string 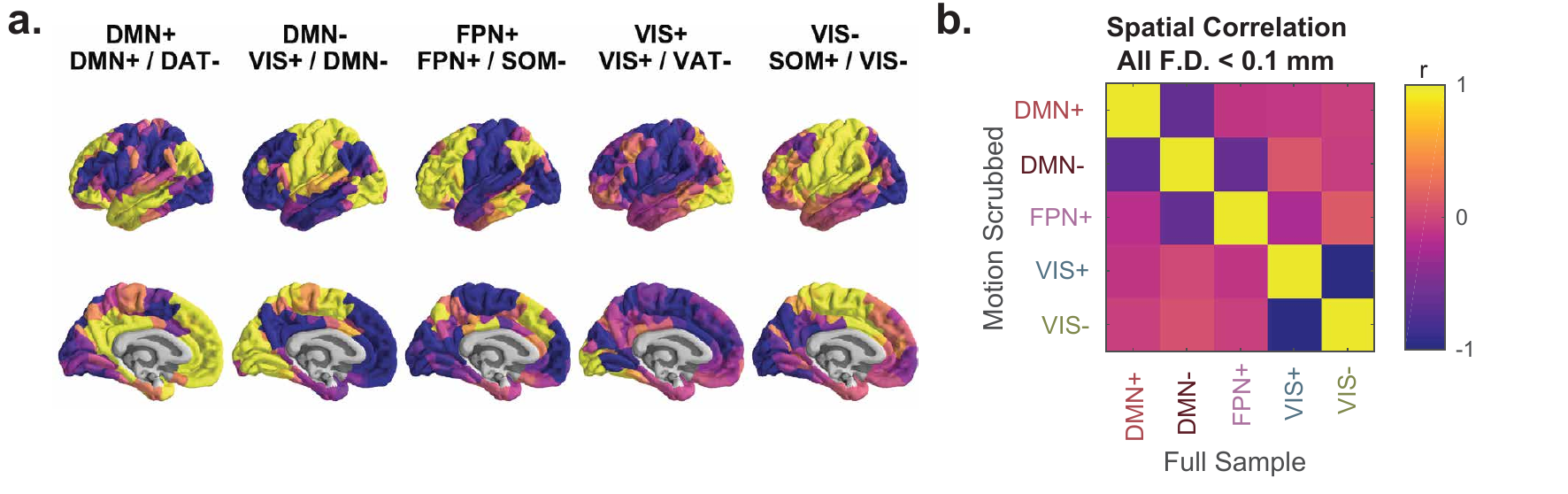}
			\caption{\textbf{Brain states after removing high motion frames.}} \emph{(a)} Cluster centroids at $k=5$ after removing 76,339 volumes with $>0.1$mm framewise displacement leaving a total of $N=226916$ volumes. \emph{(b)} Spatial correlation between motion-scrubbed centroids ($y$-axis) and full sample centroids ($x$-axis).
				\label{fig:figureS10}
		\end{center}
	\end{figure*}

	Finally, we tested whether the identification of recurrent coactivation patterns with $k$-means clustering was biased by the inclusion of single frames with sub-millimeter framewise displacement. We removed 76,339 frames associated with $>0.1$ mm framewise displacement, and repeated the clustering on the remaining 226,916 frames using correlation distance and $k=5$. The resulting centroids (Fig. \ref{fig:figureS10}a) were nearly identical to the centroids found in Figure \ref{fig:figure2}a (Fig. \ref{fig:figureS10}b, all $r>0.99$ for corresponding centroids). These findings suggest that high motion frames minimally impact the clustering process. Therefore, we included these frames so that we could have the largest continuous sample of sequential frames from which to compute transition probabilities and dwell times.
		
	\subsection*{Assessing randomness, asymmetry, autocorrelation, and distance dependence of brain state sequences}
	
\begin{figure*}
		\begin{center}
			\includegraphics[width=18cm,keepaspectratio]{\string 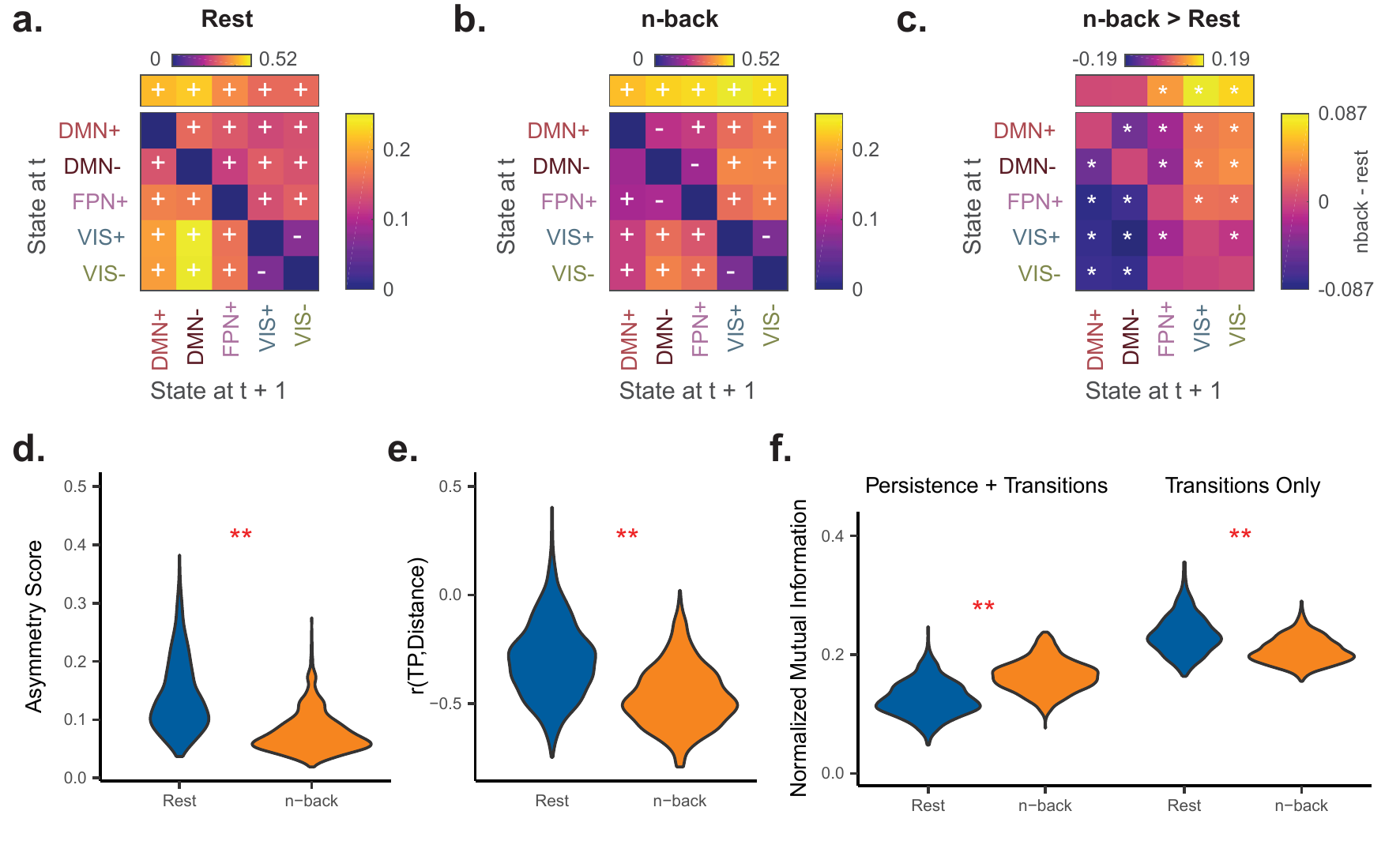}
			\caption{\textbf{Brain state transitions are context-dependent and non-random.}
				\emph{(a-b)} Group average state transition probability matrices for rest and n-back scans. Overlayed + or - indicates $p_\mathrm{corr} < 0.05$ for transitions occurring more or less, respectively, than expected under an appropriate random null model, after Bonferroni correction over 50 tests (20 transition probabilities and 5 persistence probabilities for rest and n-back). Persistence probabilities are removed from the diagonal and depicted above the transition matrix. \emph{(c)} Non-parametric permutation testing demonstrating differences between the rest and n-back group average transition probability matrices. *, $p_\mathrm{corr} < 0.05$, after Bonferroni correction over 25 tests: 20 transition probabilities and 5 persistence probabilities. \emph{(d)} Subject-level distributions of matrix asymmetry scores demonstrate that resting state transition probabilities are asymmetric relative to n-back. \emph{(e)} Subject-level distributions of the correlation between transition probabilities and Euclidean distance between states for rest (left) and n-back (right). \emph{(f)} Single-frame lagged, normalized auto mutual information for rest and n-back computed with full state time series \emph{(left)} or transition sequence only \emph{(right)}. **, $p < 10^{-15}$. Paired $t$-tests were used in panels \emph{(d-f)}. \emph{TP}, transition probability.
				\label{fig:figureS13_5}}
		\end{center}
	\end{figure*}

	In addition to assessing the relationship between brain state transitions, cognitive demands, and behavior, we were also interested to assess important basic properties of the state transition probability matrix. First, we were interested to validate previous findings which suggest that the brain does not undergo every possible transition with equal probability \cite{Vidaurre2017}. For these analyses, we begin with a transition matrix (Fig. \ref{fig:figureS13_5}a-b) whose $ij^\mathrm{th}$ element indicates the probability that state $i$ occurs at time $t$ and state $j$ occurs at time $t + t_r$, where $t_r$ is the repetition time (TR) of the scan (here, 3 seconds). The state transition probability matrix houses several pieces of important information. We refer to the diagonal entries in the transition probability matrix as the persistence probabilities, because they indicate the probability of remaining in a given state, and we refer to the off-diagonal entries in the transition probability matrix as the transition probabilities, because they indicate the probability of transitioning between two distinct states. Given this structure, we were interested to determine whether the brain dynamics that we observed could occur in a uniformly random distribution of states and state transitions. To test the randomness of persistence probabilities, we generated subject-level null state time series that preserved fractional occupancy but shuffled the temporal sequence of states; for example, if the state time series was given by the vector [1 1 2 2 3 3], then we would permute the order of the entries in that vector uniformly at random, yielding a distribution of vectors with the same proportion of each state, i.e. [1 2 1 3 3 2]. By comparing the observed persistence probabilities to the persistence probabilities in this null model, we can test whether the observed persistence probabilities would be expected based solely on fractional occupancy. We averaged together all subjects to generate a distribution of group average null persistence probabilities and compared them to the empirically observed group average persistence probabilities. 
	
	To test the randomness of transition probabilities, we generated null state time series that preserved only the states involved in transitions and reduced sequences of repeating states to a single state. We removed repeating states to control for the potentially independent effects of state persistence, which is equivalent to temporal autocorrelation, in estimating transition probabilities. For example, if the state time series was given by the vector $V = [1 1 2 2 3 3]$, then we would reduce that original vector to the new vector $V_t = [1 2 3]$, and subsequently permute the new vector uniformly at random. Specifically, for subject $i = 1,..,..N$, we reduce the state sequence vector $V_i$ to a transition sequence vector $V_{t_i}$ by eliminating repeating states, compute a transition matrix $T_i$, and average across all subjects to generate a group average transition matrix $T=\frac{1}{N}\sum_{i=1}^{N} T_i$ that excludes state persistence (i.e. diagonal entries of $T_i$ are equal to 0). Next, we shuffle $V_{t_i}$ uniformly at random to generate $V_{n_i}$, compute a transition matrix $T_{n_i}$, and average across all subjects to generate a group average null transition matrix $T_n=\frac{1}{N}\sum_{i=1}^{N} T_{n_i}$ that excludes state persistence. We generate a distribution of $T_n$ by independently shuffling $V_{t_i}$ for each subject many times and averaging them across subjects. Finally, we compare each element of $T$ to the corresponding element in a distribution of $T_n$ to compute a two-tailed, non-parametric $p$-value for each transition. These analyses demonstrated that almost all of the observed persistence and transition probabilities in both resting state and the n-back task were unexpected under these uniformly random null models (all $p_\mathrm{corr} < 0.05$ except for DMN- to DMN+ during n-back, Fig. \ref{fig:figureS13_5}a-b).
	
	Next, we assessed the properties of these transition matrices, such as the matrix symmetry, which reflects whether transitions from state 1 $\rightarrow$ 2 occur as frequently as transitions from state 2 $\rightarrow$ 1, and so forth for every state pair. Specifically, we quantified the asymmetry $\psi$ of a $k$-by-$k$ transition matrix $\mathbf{A}$ as: 
	$$\psi = 0.5 \times \frac{\sum\limits_{i=1}^k \sum\limits_{j=1, j \neq i}^k \lvert \mathbf{A}_{ij} - \mathbf{A}_{ij}^{\intercal} \rvert}{\sum\limits_{i=1}^k \sum\limits_{j=1, j \neq i}^k \lvert \mathbf{A}_{ij} \rvert}~. $$
	
	\noindent In calculating this asymmetry score, we exclude the elements along the diagonal of $\mathbf{A}$ so as to only capture directional bias in transitions between pairs of states, without including the probability of persisting in each state. The values of this score range from 0 to 1, where 0 represents a matrix that is symmetric about the diagonal and 1 represents a matrix in which the upper triangle is $-1 \times$ the lower triangle.
	
	We were also interested in how much information about future states was contained within the current state. Drawing from information theoretic approaches to analysis of discrete signals, we computed the normalized auto mutual information (NMI) between lagged state time series to answer this question. Here, we asked whether the current state contains information about the subsequent state by computing NMI between the original state time series and a state time series lagged by one element. First, we created two copies of the state time series. Then, we removed the first element from one, $\mathbf{X}$, and the last element from the other, $\mathbf{Y}$, to generate two vectors of equal length such that $\mathbf{X}_i = \mathbf{Y}_{i+1}$. We computed the NMI between $\mathbf{X}$ and $\mathbf{Y}$ as:
	$$  \frac{\mathbf{H}(\mathbf{X}) - \mathbf{H}(\mathbf{X} | \mathbf{Y})}{\mathbf{H}(\mathbf{X})}~, $$
	
	\noindent where $\mathbf{H}(\mathbf{X})$ is the entropy of $\mathbf{X}$ and $\mathbf{H}(\mathbf{X}|\mathbf{Y})$ is the conditional entropy of $\mathbf{X}$ given $\mathbf{Y}$. Where $k$ is the number of brain states, $$\mathbf{H}(\mathbf{X}) = \sum\limits_{i=1}^k P(\mathbf{X} = i) \times log(P(\mathbf{X} = i))~,$$ and $$\mathbf{H}(\mathbf{X}|\mathbf{Y}) = \sum\limits_{i=1}^k \sum\limits_{j=1}^k P(\mathbf{X} = i \land \mathbf{Y} = j) \times log(P(\mathbf{X} = i \land \mathbf{Y} = j))~.$$ In normalizing by $\mathbf{H}(\mathbf{X})$, we ensure that the NMI ranges from 0 to 1, with 0 representing two completely independent signals and 1 representing two identical signals.
	
	We anticipated that stimulus driven activity in the n-back task that occurs independently of the current brain state would result in a reduction of directional, asymmetric state transitions relative to resting state and a reduction in the dependency of state transitions on the current state. Indeed, using the normalized measure of matrix skewness described above, we found that transition probabilities at rest were more asymmetric than during the n-back task (Fig. \ref{fig:figureS13_5}d; paired $t$-test, $\mu_{nback - rest} = -0.062$, $df=878$, $t=-24.96$, $p < 10^{-15}$)s. Additionally, using the normalized auto-mutual information metric described above, we found that the current brain state carried less information about the subsequent state during the n-back task relative to rest, even when controlling for autocorrelation (Fig. \ref{fig:figureS13_5}f: normalized auto mutual information, $\mu_{nback - rest} = -0.028$, $df=878$, $t=-20.88$, $p_\mathrm{corr} < 10^{-15}$). Consistent with our hypothesis, we also found that the Euclidean distance between states was anticorrelated with the transition probabilities between states (Fig. \ref{fig:figureS13_5}e). Interestingly, however, the effect was stronger for n-back than for rest (Fig. \ref{fig:figureS13_5}e; paired $t$-test, $\mu_{nback - rest} = -0.17$, $df=878$, $t=-22.74$, $p < 10^{-15}$), suggesting that the brain is more prone to transition between distant states while at rest. 
	
	\subsection*{Transition probabilities within task blocks}
	
\begin{figure*}
		\begin{center}
			\includegraphics[width=18cm,keepaspectratio]{\string 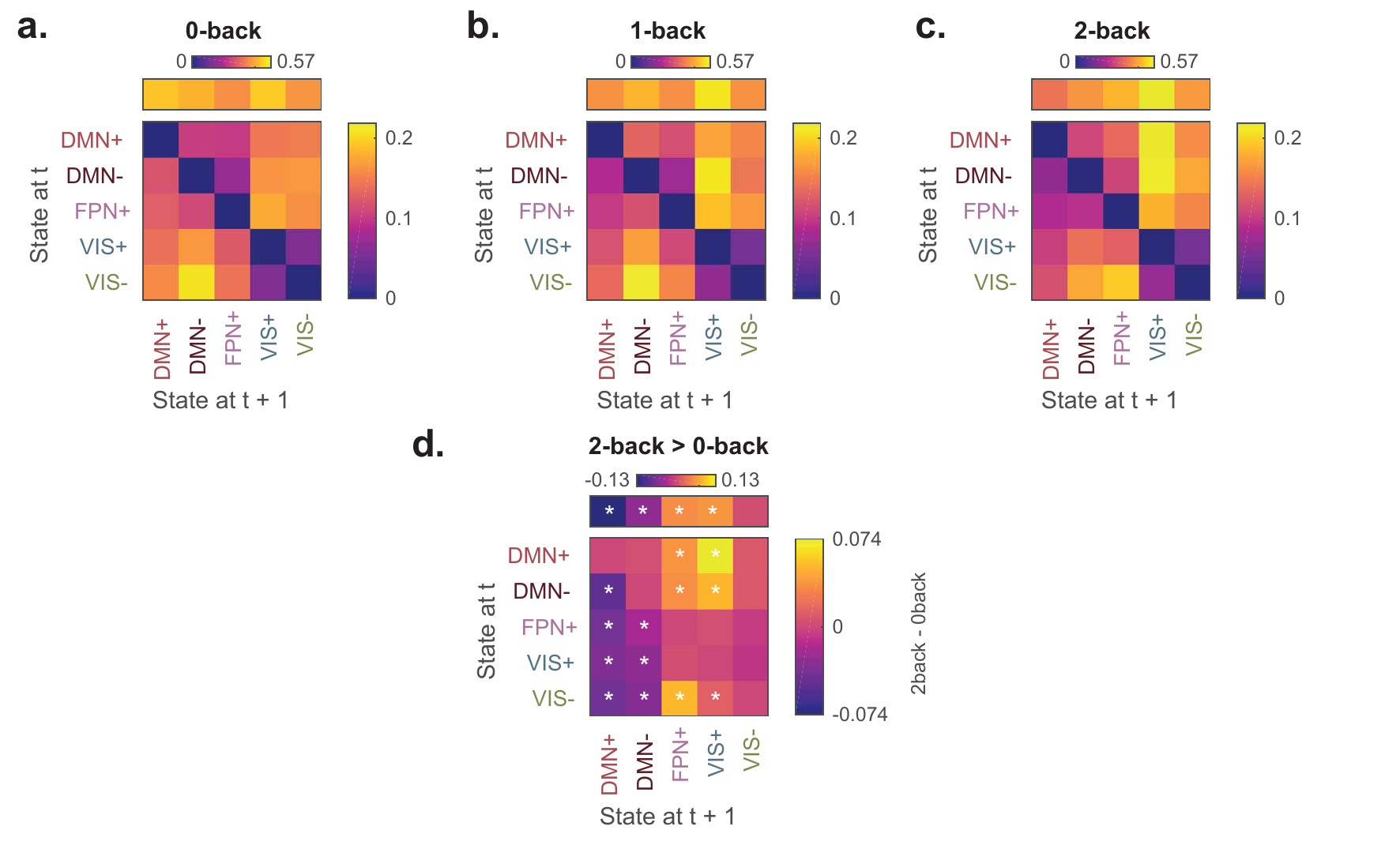}
			\caption{\textbf{Brain state transitions within task blocks.} \emph{(a-c)} Group average state transition probability matrices for 0-back (panel \emph{a}), 1-back (panel \emph{b}), and 2-back (panel \emph{c}). Persistence probabilities are removed from the diagonal and depicted above the transition matrix. Heatmap color scale represents the probability of transitioning between two states from one BOLD frame to the next, or persisting in the same state. \emph{(d)} Non-parametric permutation testing comparing transition probabilities in the 2-back to the 0-back block. *, two-tailed non-parametric $p_\mathrm{corr}<0.05$, after Bonferroni correction over 20 transition probabilities and 5 persistence probabilities.
					\label{fig:figureS14}}
		\end{center}
	\end{figure*}

	In addition to computing transition probabilities across the entire n-back task scan, we also computed transition probabilities within each task block. Because the instances of a specific task block (i.e. 0-back, 1-back, 2-back) are not continuous, we counted the number of each transition found within all instances of a given task block, and then we divided the counts by the total number of possible transitions within all instances of that task block (Fig. \ref{fig:figureS14}a-c). Similar to the analysis of transition probabilities in the entire n-back scan, we found that in the 2-back block, transitions from the DMN+ and DMN- states into the VIS+ state were increased relative to the 0-back block (Fig. \ref{fig:figureS14}d). Interestingly, we saw that transitions from DMN+ and DMN- states in to FPN+ states increased from 0-back to 2-back (Fig. \ref{fig:figureS14}d), although in the resting state these transitions were more frequent relative to the entire n-back scan (Fig. \ref{fig:figure3}c). However, transitions from VIS+ to FPN+ did not differ between the two conditions, while transitions into DMN+ and DMN- states decreased (Fig. \ref{fig:figureS14}d). These findings suggest that increasing cognitive load biases the traversal of some trajectories in state space without affecting others.
	
	\subsection*{Spatially embedded null models}
	
	To assay the specificity of brain state activity patterns themselves, we compared $\mathbf{P}_e$ for actual brain states to $\mathbf{P}_e$ for a distribution of null brain states. We generated null states using a recent method developed to find overlap between activation maps while accounting for spatial clustering of activity \cite{Alexander-Bloch2018}. Following this method \cite{Alexander-Bloch2018}, we projected node-level data to a cortical surface, inflated the surface to a sphere using FreeSurfer tools, applied a rotation to the sphere, collapsed it back to a cortical surface, and extracted node-level data by averaging over vertices belonging to each region. This process preserves the spatial grouping and relative locations of regions with similar activity while still changing their absolute locations. Importantly, reflected versions of the same rotation are applied for each hemisphere, thus also preserving the symmetry of the original activity pattern.
	
	To assay the specificity of our findings to higher order topological features found in real structural brain networks, we compared $\mathbf{P}_e$ estimates to a recently developed network null model \cite{Betzel2018}, which preserves several important spatial and topological features. This model exactly preserves the degree sequence and edge weight distribution, while approximately preserving the edge length distribution and edge length-weight relationship. We also compared our findings to a more commonly used topological null model \cite{Rubinov2010} which preserves only the degree distribution, but not the degree sequence, of the network. 
	
	\subsection*{Comparing transition energies and transition probabilities using control theory}
	
	A main goal of the present work was to provide a mechanistic description for how the brain's large-scale white matter architecture constrains its progression through activation space. To accomplish this goal, we began with a simple model of linear, time-invariant dynamics along the white matter structural connectome estimated from diffusion tractography. We represent the volume-normalized, fractional anisotropy-weighted structural network as the graph $\mathcal{G} = (\mathcal{V}, \mathcal{E})$, where $\mathcal{V}$ and $\mathcal{E}$ are the vertex and edge sets, respectively. Let $\mathbf{A}_{ij}$ be the weight associated with the edge $(i, j) \in \mathcal{E}$, and define the weighted adjacency matrix of $\mathcal{G}$ as $\mathbf{A} = [\mathbf{A}_{ij}]$, where $\mathbf{A}_{ij} = 0$ whenever $(i, j)\notin\mathcal{E}$. We associate a real value with each of the $N$ brain regions to generate a vector describing the activity in each node at time $t$, and we define the map $\mathbf{x} : \mathbb{R}_{\geq0} \rightarrow \mathbb{R}^{N}$ to describe the dynamics of activity in network nodes over time. Here we employ a simplified noise-free linear continuous-time and time-invariant model of such dynamics:
	
	\begin{align}
		\dot{\mathbf{x}}(t) = \mathbf{Ax}(t) + \mathbf{B}\mathbf{u}(t)~,
		\end{align}\label{eq:cont2}
	
	\noindent where $\mathbf{x}$ describes the activity (i.e. voltage, firing rate, BOLD signal) of brain regions over time. Thus, the vector $\mathbf{x}$ has length $N$, where $N$ is the number of brain regions in the parcellation, and the value of $\mathbf{x}_i$ describes the activity level of that region. The matrix $\mathbf{A}$ is symmetric, with the diagonal elements satisfying $\mathbf{A}_{ii} = 0$. Prior to calculating control energy, we divide $\mathbf{A}$ by $\xi_0(\mathbf{A})$ and subtract 1 from the diagonal elements of $\mathbf{A}$, where $\xi_0(\mathbf{A})$ is the largest eigenvalue of $\mathbf{A}$. This step makes the system marginally stable by ensuring that the maximum eigenvalue of the system is equal to 0. The input matrix $\mathbf{B}$ identifies the control input weights, which we set to the $N\times N$ identity matrix by default. For certain analyses (Fig. \ref{fig:figure5}d, Fig. \ref{fig:figureS20}b and d, Fig. \ref{fig:figureS8}d, Fig. \ref{fig:figureS7}d), a set of brain regions $\mathcal{K} \subset \mathcal{V}$ that belong to a particular cognitive system \cite{ThomasYeo2011} was given increased weight, such that
	
	\begin{align}
	\mathbf{B}_{ii} = 
	\begin{cases} 
		c & \mathrm{if}\: i \in \mathcal{K} \\ 
		1 & \mathrm{otherwise}
	\end{cases}
	,
	\end{align}
	
	\noindent and $c$ is a positive, real scalar value, which we set equal to 2 here. The input $\mathbf{u} : \mathbb{R}_{\geq0}\rightarrow\mathbb{R}^M$ denotes the control strategy.
	
	To compute the minimum control energy required to drive the system from an initial activity pattern $\mathbf{x}_o$ to a final activity pattern $\mathbf{x}_f$ over some time $T$, we compute an invertible controllability Gramian $\mathbf{W}$ for controlling the network $\mathbf{A}$ from the set of network nodes $\mathcal{K}$ (in our case, every node in the network), where:
	
		\begin{align}
		\mathbf{W} = \int_{0}^{T}e^{\mathbf{A}(T-\tau)}\mathbf{B}\mathbf{B}^\intercal e^{\mathbf{A^\intercal}(T-\tau)}d\tau,
		\end{align}
	
	\noindent where $T$ is the time horizon, which specifies the time over which input to the system is allowed. After computing the controllability Gramian, we can solve for the minimum control energy $\mathbf{E_m}$ by computing the quadratic product between the inverted controllability Gramian and the difference between $\mathbf{x}_o$ and $\mathbf{x}_T$:
	\begin{align}
	\mathbf{E_m} = (e^{\mathbf{A}T}\mathbf{x_o}-\mathbf{x_T})^{\intercal}\mathbf{W}^{-1}(e^{\mathbf{A}T}\mathbf{x_o}-\mathbf{x_T})~.
	\end{align}
	
	\noindent In Fig. \ref{fig:figure5}b, we computed the $k\times k$ transition energy matrix $\mathbf{T}_e$ as the minimum energy required to transition between all possible pairs of the $k$ clustered brain states, given the white matter connections represented in $\mathbf{A}$. We refer to the on-diagonal elements of $\mathbf{T}_e$ as \textit{persistence energies}, because they quantify the energy required to maintain $\mathbf{x}_o$ in the special case where $\mathbf{x}_o = \mathbf{x}_T$. We refer to the off-diagonal elements of $\mathbf{T}_e$ as \textit{transition energies}, because they quantify the energy required to move between all pairs of $\mathbf{x}_o$ and $\mathbf{x}_T$ where $\mathbf{x}_o \neq \mathbf{x}_T$.
	
\begin{figure*}
		\begin{center}
			\includegraphics[width=18cm,keepaspectratio]{\string 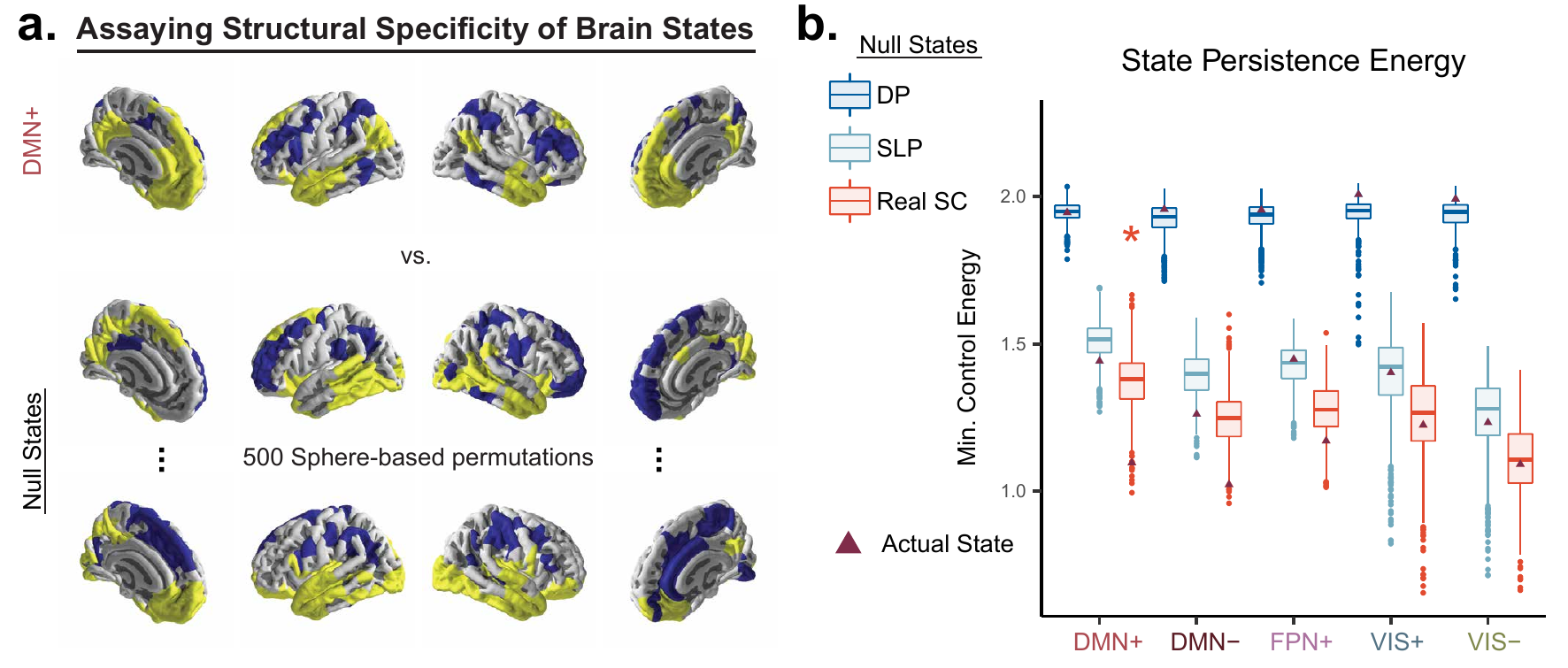}
			\caption{\textbf{Spatial and topological properties of brain structure facilitate selective brain state stability.} \emph{(a)} Construction of null states preserving symmetry and spatial clustering of activity using sphere-based permutation \cite{Alexander-Bloch2018}. We compare the minimum control energy required to maintain the brain in each state (persistence energy) relative to spatially permuted states. \emph{(b)} We computed persistence energy for each state and its permuted variants in group average SC (orange) and found that the DMN+ state required less energy to persist than its permuted variants. We performed the same test in two null models: a null model preserving topology (blue) and a null model preserving both topology and spatial constraints (light blue). Orange *, $p_\mathrm{corr} < 0.05$ after Bonferroni correction over each of the 5 states, separately for each null model. \emph{(Deg. Pres.)}, degree distribution-preserving null model \cite{Rubinov2010}. \emph{Space Pres.}, degree sequence, edge weight and length distribution and relationship preserving null model \cite{Betzel2018}.
				\label{fig:figureS19}}
		\end{center}
	\end{figure*}

\begin{figure*}
		\begin{center}
			\includegraphics[width=18cm,keepaspectratio]{\string 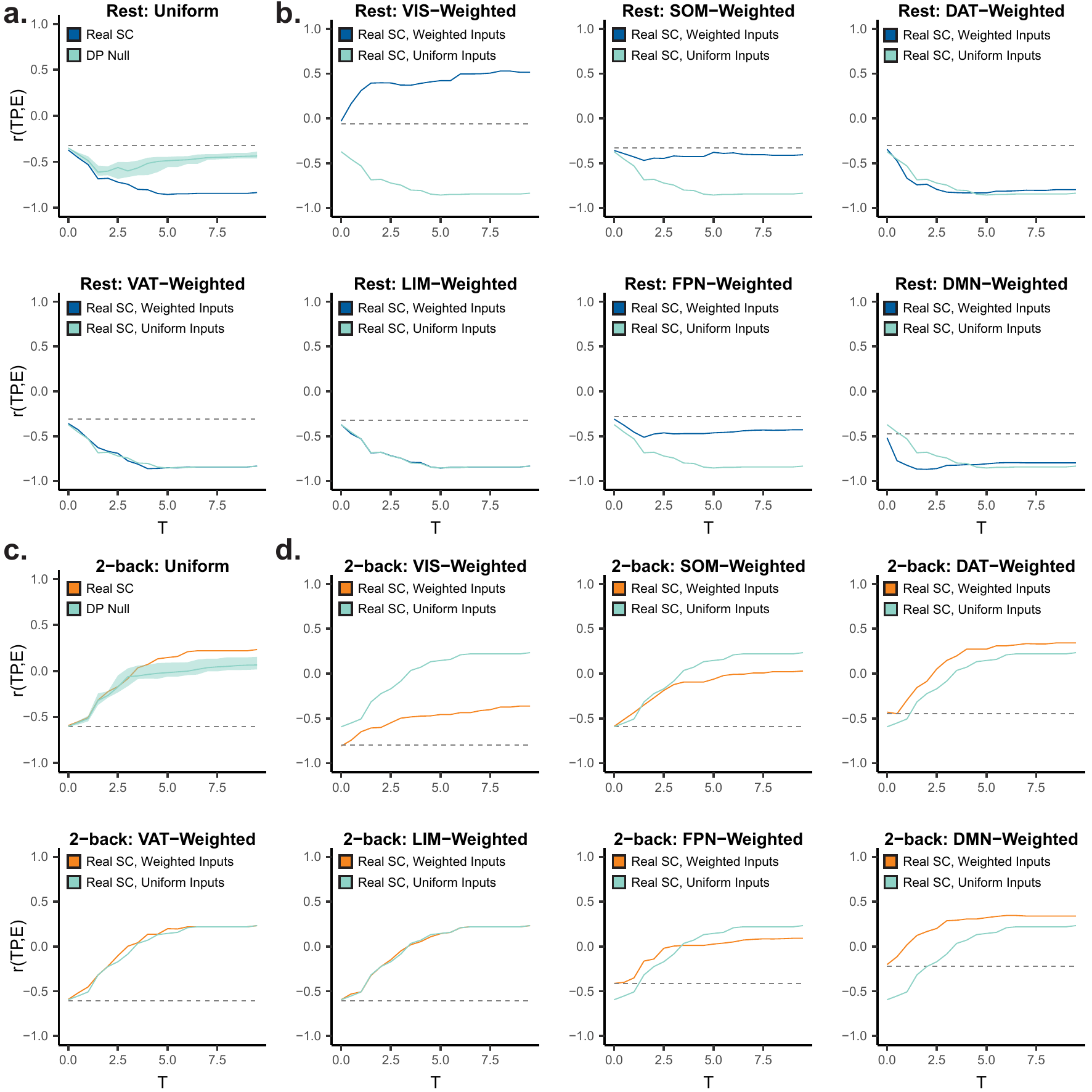}
			\caption{\textbf{Selection of control horizon $T$ and weighted control.} \emph{(a-d)} Relationship between control horizon $T$ ($x$-axis) and Spearman rank correlation ($y$-axis) between structurally predicted transition energies and empirically observed transition probabilities for resting state data (panel \emph{a-b}) and 2-back task data (panel \emph{c-d}). \emph{(a, c)} Transition energies are computed for uniformly weighted inputs. Dashed line represents correlation between transition probabilities and $||\mathbf{x}_T- \mathbf{x}_o||$. Teal trace represents Spearman correlation using transition energies obtained from a distribution of null networks with preserved degree sequence (DP Null). Shaded region represents the full range of this distribution. \emph{(b, d)} Transition energies are computed by weighting inputs towards different cognitive systems \cite{ThomasYeo2011}. Dashed line represents correlation between transition probabilities and weighted state-space transition distance, computed as $(\mathbf{x}_T- \mathbf{x}_o)^\intercal(\mathbf{B}_\mathcal{K}\mathbf{B}_\mathcal{K}^\intercal)^{-1} (\mathbf{x}_T- \mathbf{x}_o)$. Teal trace represents Spearman correlation using transition energies computed with uniformly weighted inputs, the exact same as the blue and orange traces in panel \emph{a} and panel \emph{c}, respectively. \emph{TP}, transition probability. \emph{E}, minimum control energy $\mathbf{E}_m$.
				\label{fig:figureS20}}
		\end{center}
	\end{figure*}

	First, we sought to determine whether the brain states that we observed (Fig. \ref{fig:figure2}a) were easier to maintain (1) compared to null states and (2) in real brain networks compared to null brain networks. Thus, we compared the persistence energy for the actual states to that of null brain states \cite{Alexander-Bloch2018} with preserved symmetry and spatial clustering (Fig. \ref{fig:figureS19}a). We found that the persistence energy of the DMN+ state was significantly lower than that of its respective null states (Fig. \ref{fig:figureS19}b: one-tailed non-parametric test, DMN+, $p_\mathrm{corr} = 0.045$, 1000 sphere-permuted null states). The DMN- state also required lower persistence energy than many of its respective null states, although this result was not significant after Bonferroni correction over all states (Fig. \ref{fig:figure5}b, one-tailed non-parametric test, $p_\mathrm{corr} = 0.05$, 1000 sphere-permuted null states). Crucially, the DP and SLP null models did not exhibit selectively increased stability in DMN states, suggesting that DMN-driven states may arise in part due to complex features of white matter topology that allow for their stability.
	
	Next, we hypothesized that the off-diagonal elements of the empirically observed state transition matrix at rest (Fig. \ref{fig:figure3}a) would be anticorrelated with the off-diagonal elements of $\mathbf{T}_e$. This hypothesis was based on the notion that the brain would empirically prefer trajectories in state space requiring smaller magnitude control inputs to achieve. Note that we ignored the diagonal of $\mathbf{T}_e$, which captures whether persistence energies explain observed state dwell times, a question that we were not sufficiently powered to ask with only $k=5$ states but that should be revisited in future studies.
	
	The choice of the control horizon $T$ is critical for the calculation of $\mathbf{T}_e$. When control inputs are uniformly distributed across the brain such that $\mathbf{B}_\mathcal{K}$ is the identity matrix, then as $T$ approaches 0, the minimum control energy $\mathbf{E}_m$ becomes proportional to the squared Euclidean norm of $\mathbf{x}_T - \mathbf{x}_o$. $||\mathbf{x}_T - \mathbf{x}_o||^2$ is the state space distance between the initial and final state in the transition. At longer time horizons, both $||\mathbf{x}_T - \mathbf{x}_o||^2$ and the topology of $\mathbf{A}$ determine $\mathbf{E}_m$. When $\mathbf{B}_\mathcal{K}$ is not the identity matrix, $\mathbf{E}_m$ becomes proportional to $(\mathbf{x}_T- \mathbf{x}_o)^\intercal(\mathbf{B}_\mathcal{K}\mathbf{B}_\mathcal{K}^\intercal)^{-1} (\mathbf{x}_T- \mathbf{x}_o)$ as $T$ approaches 0. 
	
	However, because the units of the edge weights of $\mathbf{A}$, obtained from deterministic tractography performed on diffusion-weighted imaging data, are not measured in activity per unit time, the value of $T$ is arbitrary relative to $\mathbf{A}$ and the real time in which neural activity was measured through BOLD fMRI. Therefore, we chose $T$ in a data-driven fashion by computing transition energies for a range of $T$ values using a group representative structural $\mathbf{A}$ matrix, and computing the Spearman rank correlation between transition energies and the group average transition probability matrix from resting state fMRI data (Fig. \ref{fig:figureS20}a). We used the Spearman rank correlation rather than the Pearson correlation in order to reduce the effect of outliers on estimating the relationship between transition probability and transition energy. We found that the strongest negative correlation value was obtained for $T=5$, but similarly strong negative correlations were found for the range $T=[5,10]$ (Fig. \ref{fig:figureS20}a). Accordingly, we used $T=5$ for the analyses presented in Fig. \ref{fig:figure5}b-c, Fig. \ref{fig:figure6}c, and Fig. \ref{fig:figureS19}b. We also carried out the same analysis using a distribution of null networks with preserved degree sequence (see Fig. \ref{fig:figure5}b), which revealed that there was no $T$ value yielding a correlation between transition energies and observed resting state transition probabilities as strongly negative as when we used the real structural connectivity matrix (Fig. \ref{fig:figureS20}a).
	
	In addition to controlling the brain with uniformly weighted inputs, we also asked whether transition energies obtained using a non-uniform distribution of inputs might better explain the observed transition probabilities. Specifically, we hypothesized that accounting for external visual input during the fractal n-back task might provide a more accurate estimation of the input energy needed to achieve each transition. We weighted the inputs towards one cognitive system \cite{ThomasYeo2011} at a time while still allowing input into every brain region, and then recomputed the Spearman correlation between transition energies and transition probabilities for the resting state (Fig. \ref{fig:figureS20}b) and the 2-back condition of the n-back task (Fig. \ref{fig:figureS20}d). This analysis revealed that accounting for visual input in computing transition energies improved our ability to explain the observed brain state transition probabilities during the 2-back condition (Fig. \ref{fig:figureS20}d VIS-weighted subpanel, orange trace is lower than teal trace), and abolished our ability to explain resting state transition probabilities (Fig. \ref{fig:figureS20}b VIS-weighted subpanel, blue trace is greater than 0 and dashed line is near 0). However, we did not find a clear role for brain structure in this relationship, as evidenced by the fact that transition energy did not explain transition probability any better than the weighted distance between states (Fig. \ref{fig:figureS20}d, orange trace does not dip below dashed line). Thus, we present results in Fig. \ref{fig:figure5}d for $T=0.001$, where $\mathbf{E}_m$ is proportional to $(\mathbf{x}_T- \mathbf{x}_o)^\intercal(\mathbf{B}_\mathcal{K}\mathbf{B}_\mathcal{K}^\intercal)^{-1} (\mathbf{x}_T- \mathbf{x}_o)$. Nevertheless, this finding suggests a specificity of the constraints of state-space transition distance on the brain's empirically observed progression through state space. Resting state transition probabilities can be explained by unweighted state-space distance (Fig. \ref{fig:figureS20}a, dashed line, Spearman's $r=-0.32$) but not by visual system-weighted state-space distance (Fig. \ref{fig:figureS20}b, VIS-weighted dashed line, Spearman's $r=-0.06$); 2-back transition probabilities can be explained by unweighted state-space distance (Fig. \ref{fig:figureS20}c, dashed line, Spearman's $r=-0.61$), but are explained best by visual system-weighted state-space distance (Fig. \ref{fig:figureS20}b, VIS-weighted dashed line, Spearman's $r=-0.80$). This result suggests that visual input allows the brain to deviate from the constraints of state-space distance found at rest, while an equal consideration of visual inputs alongside other inputs is key to explaining resting state transitions. Resolving the effect of structure on brain dynamics during a task may require full knowledge of all input sources, which could potentially be uncovered through a data-driven approach \cite{Ashourvan2019}. 
		
	\subsection*{Impact of parcellation and cluster number}
	
	The choice of parcellation scale may impact analyses involving tractography, where relative region sizes may bias estimates of connectivity. We chose a relatively fine-grained parcellation with 462 nodes because previous coactivation pattern analyses \cite{Chen2015} at this scale produced cluster assignments similar to those obtained by clustering at the voxel level. However, we were interested to know whether we would obtain the same results using a coarser parcellation scale. Therefore, we repeated the clustering procedure and subsequent control theoretic analyses using the Lausanne 234 node parcellation \cite{Cammoun2012}.
	
\begin{figure*}
		\begin{center}
			\includegraphics[width=18cm,keepaspectratio]{\string 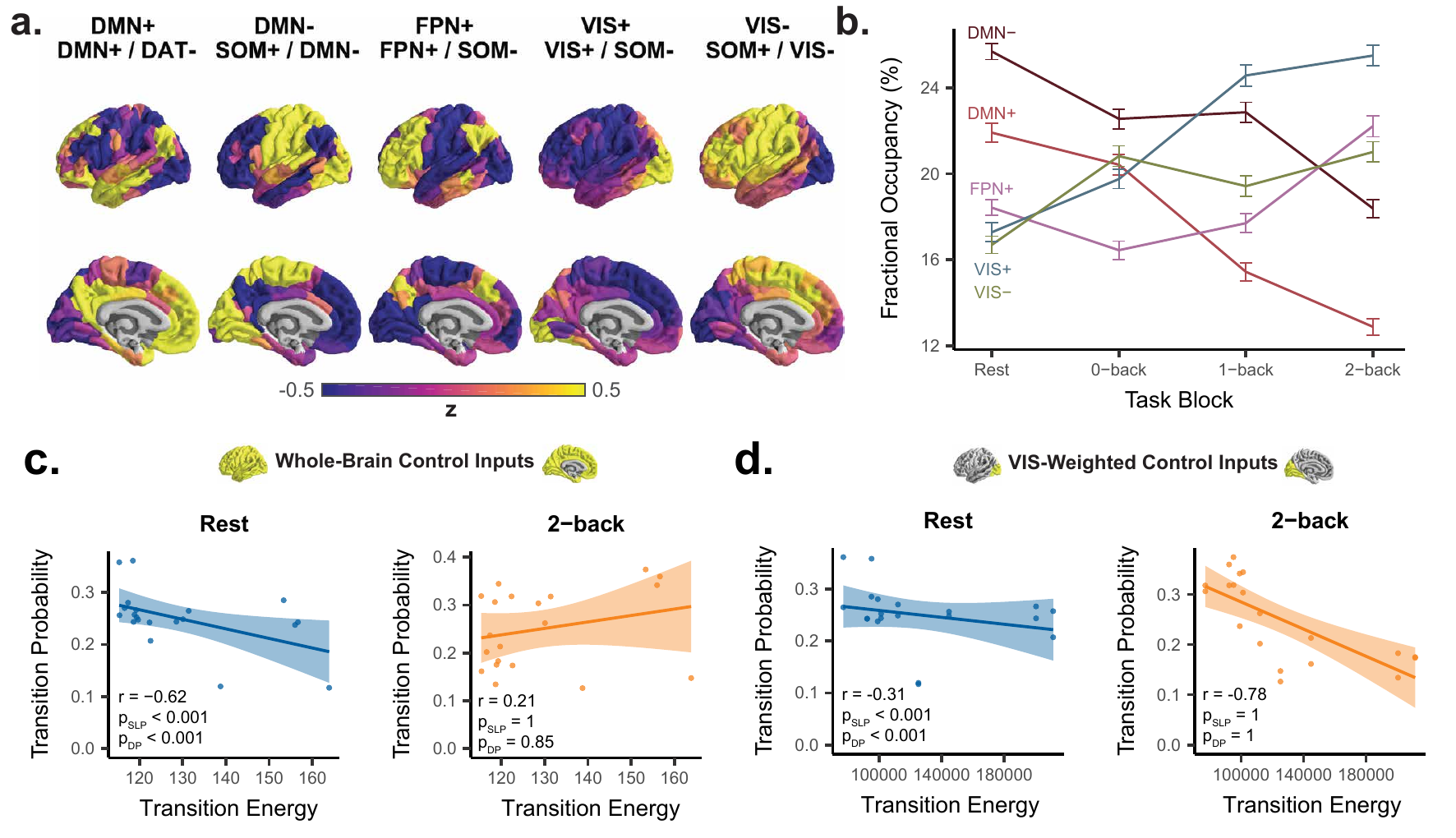}
			\caption{\textbf{Key findings reproduced at $k=5$ using the 234-node Lausanne parcellation.} \emph{(a)} Cluster centroids at $k=5$ are similar to that of the 463-node Lausanne parcellation. \emph{(b)} State fractional occupancies change with increasing cognitive load similarly compared to the 463-node parcellation analysis. \emph{(c-d)} Correlation between structure-based transition energy prediction ($x$-axis) and empirically derived transition probability ($y$-axis) for resting state (left) and the 2-back condition of the n-back task (right), using inputs weighted evenly throughout the whole brain \emph{(c)} or weighted positively towards the visual system \emph{(d)}.
				\label{fig:figureS8}}
		\end{center}
	\end{figure*}

	First, this analysis revealed brain states whose spatial maps were virtually identical to those generated using the 462 node parcellation (Fig. \ref{fig:figureS8}a). The fractional occupancy of these states also changed with cognitive load in a similar fashion (Fig. \ref{fig:figureS8}a) compared to the results in Fig. \ref{fig:figure2_5}d. Similar to the results presented in the main text, we found a negative relationship between transition energies and empirically observed brain state transition probabilities at rest (Fig. \ref{fig:figureS8}c, Spearman's $r=-0.62$, $p_{SLP} < 0.001$, $p_{DP} < 0.001$) with a weak positive relationship between transition energies and transition probabilities from the 2-back condition (Fig. \ref{fig:figureS8}c, Spearman's $r=0.21$, $p_{SLP} = 1$, $p_{DP} = 0.85$). We also found that discounting the weight of the visual system in calculating state-space transition distance allowed us to better explain 2-back transition probabilities (Fig. \ref{fig:figureS8}d, Spearman's $r=-0.78$, $p_{SLP} = 1$, $p_{DP} = 1$), and reduced our ability to explain resting state transition probabilities (Fig. \ref{fig:figureS8}d, Spearman's $r=-0.31$, $p_{SLP} < 0.001$, $p_{DP} < 0.001$). These results suggest that one can quantify the constraints of white matter architecture on brain state transitions at rest at multiple scales of region definition, supporting the generalizability of our findings.
	
	A limitation of $k$-means clustering is the need to specify an absolute number of clusters. While the data suggests that $k=5$ is the optimal solution, one could certainly choose to analyze this data at multiple scales. Because of a recent study \cite{Gutierrez-Barragan2019} in which $k=6$ was identified as the optimal solution by the same heuristic, we reproduce our results at $k=6$ in the interest of facilitating comparison between the two studies. 
	
\begin{figure*}
		\begin{center}
			\includegraphics[width=18cm,keepaspectratio]{\string 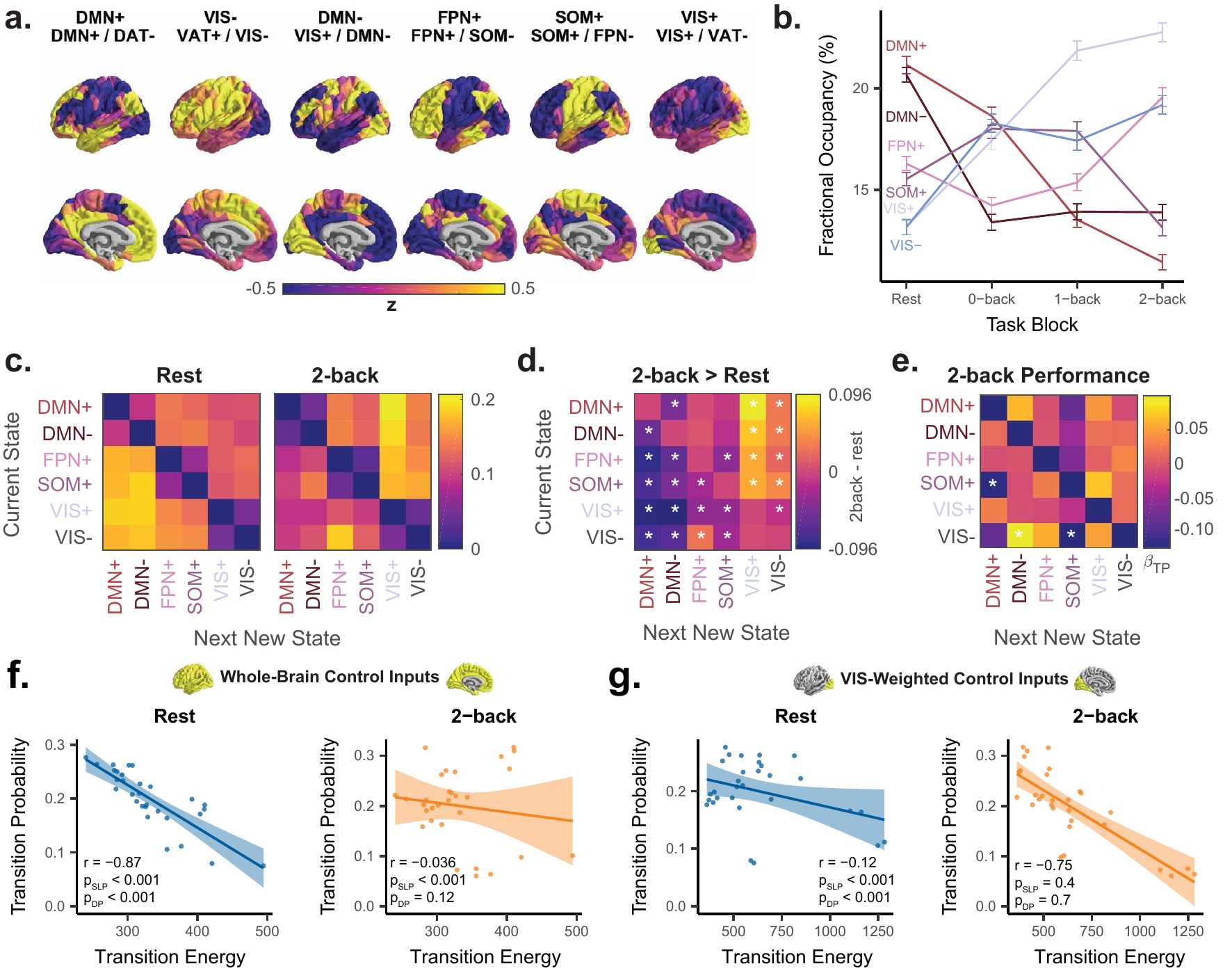}
			\caption{\textbf{Key findings reproduced at $k=6$.} \emph{(a)} Cluster centroids at $k=6$, similar to $k=5$ with the addition of a SOM+ cluster. \emph{(b)} State fractional occupancies change with increasing cognitive load similarly compared to $k=5$. \emph{(c-d)} Group average state transition probability matrices for rest (panel \emph{(c)}) and 2-back condition of the n-back task (panel \emph{(d)}) scans. \emph{(e)} Permutation testing to compare 2-back and rest transition probabilities. *, statistically significant after Bonferroni correction over 30 transitions. \emph{(f-g)} Correlation between structure-based transition energy prediction ($x$-axis) and empirically derived transition probability ($y$-axis) for resting state (left) and the 2-back condition of the n-back task (right), using inputs weighted evenly throughout the whole brain \emph{(f)} or weighted positively towards the visual system \emph{(g)}. \emph{TP}, transition probability.
				\label{fig:figureS7}}
		\end{center}
	\end{figure*}

	At $k=6$, cluster centroid brain states interestingly become grouped into 3 anticorrelated pairs (DMN+ and DMN-, VIS+ and VIS-, FPN+ and SOM+), similar to Ref. \cite{Gutierrez-Barragan2019}. The states were also highly similar to those at $k=5$, with the DMN- state from $k=5$ essentially ``split'' into the SOM+ and DMN- state at $k=6$. The DMN- state at $k=6$ state houses concurrent high amplitude activity in the dorsal attention network and visual system with low amplitude activity in the default mode network (Fig. \ref{fig:figureS7}a). This appearance of ``split'' brain states is consistent with hierarchical state organization \cite{Vidaurre2017a,Chen2018}, which becomes apparent at different clustering scales. As WM load increased from 0-back to 2-back, we saw a decrease in DMN+ fractional occupancies with a concurrent increase in FPN+ and VIS state fractional occupancies (Fig. \ref{fig:figureS7}b). Direct transitions between anticorrelated states were infrequent (Fig. \ref{fig:figureS7}c-d), with a drastic shift in transition probabilities towards VIS states and away from DMN states during 2-back relative to rest (Fig. \ref{fig:figureS7}e). Similar to $k=5$, at $k=6$ we found that 2-back task performance was related to 2-back state transitions from the VIS- state into states with coherent frontoparietal and default mode activity (Fig. \ref{fig:figureS7}e). Specifically, we found that transitions from VIS- into DMN- and FPN+ were positively associated with performance, consistent with the high amplitude dorsal attention network activity and low amplitude DMN activity found in the DMN- state (Fig. \ref{fig:figureS7}a). Additionally, transitions from the VIS- state to the SOM+ state were negatively associated with performance, consistent with the frontoparietal deactivation found in the SOM+ state (Fig. \ref{fig:figureS7}a). We again found a strong negative correlation between transition energies and resting state transition probabilities (Fig. \ref{fig:figureS7}f, Spearman's $r=-0.87$, $p_{SLP} < 0.001$, $p_{DP} < 0.001$) that was specific to the topology of white matter, supporting the notion that the constraints of white matter on state-space progression generalizes across multiple scales of states. Similar to our results at $k=5$, we found that discounting the weight of the visual system in calculating the state-space distance of transitions allowed us to better explain the state transitions observed during the 2-back task, during which visual stimuli are frequently delivered (Fig. \ref{fig:figureS7}g, Spearman's $r=-0.75$, $p_{SLP} = 0.4$, $p_{DP} = 0.7$). Overall, this analysis suggests that our main findings hold true at near optimal values of $k$, and that much can be learned from studying the states of the brain at different scales.
	
	\begin{figure*}
		\begin{center}
			\includegraphics[width=18cm,keepaspectratio]{\string 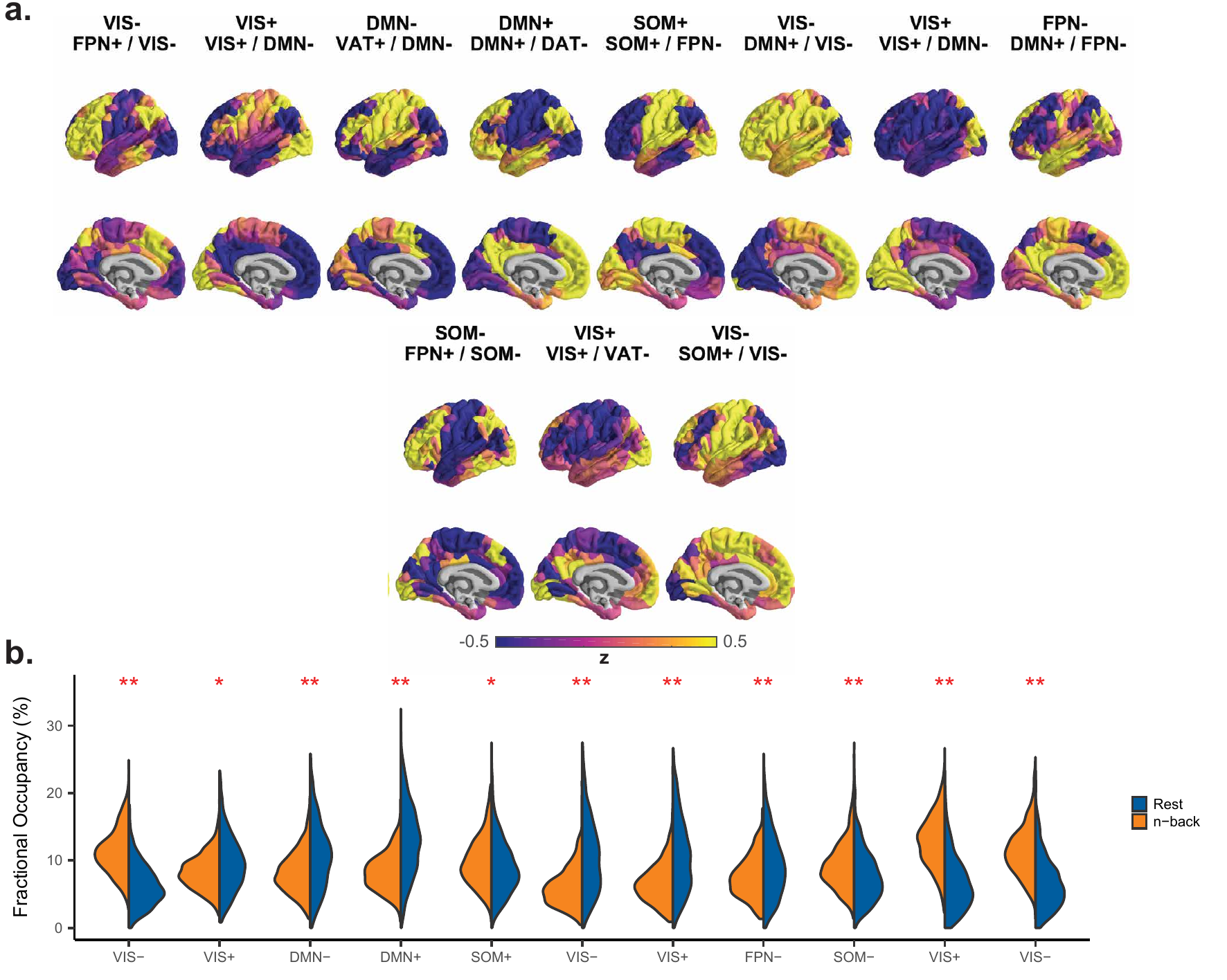}
			\caption{\textbf{Brain states at $k=11$.} \emph{(a)} Cluster centroids at $k=11$ are reminiscent of brain states at $k=5$ but with several additional combinations of resting state network activity patterns. \emph{(b)} Nearly every state at $k=11$ has a different fractional occupancy for rest and n-back. *, $p<10^{-4}$, **, $p < 10^{-15}$.
				\label{fig:figureS9}}
		\end{center}
	\end{figure*}

\end{document}